\pdfoutput=1 

\documentclass{amsart}
\usepackage{mathrsfs}
\usepackage{amssymb}
\usepackage{graphicx}

\usepackage{color}

\theoremstyle{definition}

\newtheorem{example}{Example}

\theoremstyle{remark}

\numberwithin{equation}{section}

\def\Xint#1{\mathchoice
{\XXint\displaystyle\textstyle{#1}}%
{\XXint\textstyle\scriptstyle{#1}}%
{\XXint\scriptstyle\scriptscriptstyle{#1}}%
{\XXint\scriptscriptstyle\scriptscriptstyle{#1}}%
\!\int}
\def\XXint#1#2#3{{\setbox0=\hbox{$#1{#2#3}{\int}$}
\vcenter{\hbox{$#2#3$}}\kern-.5\wd0}}

\def\dashint{\Xint-}

\def\d{\text{d}}
\def\e{\text{e}}
\def\i{\text{i}}

\begin{document}

\title[Products of random matrices and quantum point scatterers]
  {Products of random matrices and generalised quantum point scatterers} 

\author{Alain Comtet}
\address{Laboratoire de Physique Th\'eorique et Mod\`eles
  Statistiques, B\^at. 100, Universit\'e Paris-Sud, UMR 8626 du CNRS, 
  F-91405 Orsay Cedex, France~;
  Universit\'e Pierre et Marie Curie -- Paris 6,
  4 place Jussieu,
  F-75552 Paris Cedex, France}  
  \email{alain.comtet@u-psud.fr}        

\author{Christophe Texier}
\address{Laboratoire de Physique Th\'eorique et Mod\`eles
  Statistiques, B\^at. 100, Universit\'e Paris-Sud, UMR 8626 du CNRS, 
  F-91405 Orsay Cedex, France~;   
  Laboratoire de Physique des Solides, B\^at. 510, Universite
  Paris-Sud, UMR 8502 du CNRS, F-91405 Orsay Cedex, France}  
  \email{christophe.texier@u-psud.fr}        

\author{Yves Tourigny}
\address{School of Mathematics,
        University of Bristol,
        Bristol BS8 1TW, United Kingdom}
  \email{y.tourigny@bristol.ac.uk}

\thanks{We thank Jean--Marc Luck for drawing to our attention the work
  of T. M. Nieuwenhuizen, and Tom Bienaim\'e for participating in
  the study of the supersymmetric model.\\
  DOI : 10.1007/s10955-010-0005-x. 
  The final publication is available at www.springerlink.com}


\date{June 14, 2010}


\begin{abstract}
To every product of $2\times2$ matrices, there
corresponds a one-dimensional Schr\"{o}dinger equation whose potential
consists of generalised
point scatterers. 
Products of {\em random}
matrices are obtained by making these interactions and their
positions random.
We exhibit a simple one-dimensional quantum model
corresponding to the most general product of matrices in 
$\text{SL}\left( 2, {\mathbb R} \right)$.
We use this correspondence to find new examples of products of random
matrices for which the invariant measure can be expressed in simple
analytical terms.
\end{abstract} 

\maketitle

\section{Introduction}
\label{introductionSection}
Products of random $2 \times 2$ matrices arise in many physical contexts:
in the study of random spin chains, or when calculating the distribution of the
natural frequencies of a classical random spring chain, or  more generally when
considering the propagation of a wave in a one-dimensional disordered
medium \cite{BL,LGP,Lu}.
It is often the case that, in the presence of disorder
(i.e. randomness), the waves become sharply localised in space. 
This physical phenomenon is known as {\em Anderson localisation}; one of its
mathematical manifestations is the exponential
growth of the product of random matrices. 

The rate of growth is called the {\em Lyapunov exponent}; it often has
a physical interpretation 
in terms of the exponential decay of the transmission probability
as the size of the disordered region grows.
One method for calculating the Lyapunov exponent is based on a general
theory developed by Furstenberg and others \cite{BL,CaLa,Fu}. This
method 
requires the explicit knowledge of a certain  measure on the
projective space, invariant under the action of the matrices in the
product. 
Examples of products of random matrices for which this invariant
measure can be obtained in 
analytical form are, however, very few; see for instance
\cite{BL,ChLe,Ish73,MTW} and the references therein. 

The calculation of the Lyapunov exponent
need not always make use of this invariant
measure. There are alternative approaches; see for instance
\cite{Lu,Ni,Pol10}.
Nevertheless, the problem of determining the invariant measure is
interesting in itself, 
and the present paper will focus on the presentation of new explicit examples
from corresponding examples of exactly solvable
models of one-dimensional disordered systems with {\em point
  scatterers} \cite{ADK,AGHH,Se}. 
In our context, the phrase ``exactly solvable'' means that the
calculation of the 
Lyapunov exponent associated with the disordered system is reduced to
a problem of quadrature. 
Some of the models were solved by Nieuwenhuizen \cite{Ni} (without the
use of the invariant measure); some of
them are, apparently, new. 
Although the work reported here is, for the most part, mathematically
driven, these new 
models are of independent physical interest.
To the best of our knowledge, all the explicit formulae for the
invariant measures constitute new results.
In the remainder of this introductory section, we review some relevant
concepts and some known 
facts, summarise our main results, and give
a sketch of our approach.

\subsection{Products of random matrices}
\label{productSubsection}

Let 
$$
A_1,\, A_2,\, A_3,\,\cdots
$$
denote independent, identically-distributed $2\times2$ matrices with
unit determinant, let $\mu$ be their common distribution, and 
consider
the product
\begin{equation}
\Pi_n :=  A_{n} A_{n-1} \cdots A_1\,.
\label{productOfMatrices}
\end{equation}
The number
\begin{equation}
  \gamma_\mu
  := \lim_{n \rightarrow \infty} 
  \frac{{\mathbb E} \left ( \ln \left | \Pi_n \right | \right )}{n}
  \label{lyapunovExponent}
\end{equation}
where $|\cdot|$ denotes the norm on matrices induced by the euclidean
norm on vectors, also denoted $|\cdot|$, 
is called the {\em Lyapunov exponent} of the product.

The product grows if the angle between the columns decreases or, equivalently, if the columns tend to align
along some common direction.
In precise mathematical terms, a {\em direction} in ${\mathbb R}^d$ is a straight line through the origin, and the set
of all directions is, by definition, the projective space $P \left ( {\mathbb R}^d \right )$. The case $d=2$ is particularly simple:
any direction
$$
\left \{ \lambda \begin{pmatrix}
x \\
y
\end{pmatrix} \,: \;\lambda \in {\mathbb R} \right \} \,,
$$
is characterised by
the reciprocal, say 
$$
z =  \frac{x}{y} \in \overline{\mathbb R} := {\mathbb R} \cup \{\infty\}\,,
$$
of its slope. So we can identify $P\left( {\mathbb R}^2 \right)$ with
$\overline{\mathbb R}$. 
The calculation of the Lyapunov exponent is often based on the formula
\cite{BL,CaLa,Fu}: 
\begin{equation}
  \gamma_\mu = 
  \int_{\overline{\mathbb R}} \nu ( \d z )
   \int_{\text{SL}\left(2,{\mathbb R} \right)}\mu (\d A) \, 
  \ln \frac{\left | A \begin{pmatrix} z \\ 1 \end{pmatrix} \right |}{ \left |  
  \begin{pmatrix} z \\ 1 \end{pmatrix} \right |} \, .
\label{furstenbergFormula}
\end{equation}
In this expression, $\mu$ is the {\em known} common distribution of
the matrices $A_n$ in the product, whereas 
$\nu$ is the--- a priori {\em unknown}---probability measure on the
projective line 
which is invariant under the action of matrices drawn from 
$\mu$. Here, invariance means that if 
$$
A = \begin{pmatrix}
a & b \\
c & d
\end{pmatrix}
$$
is a $\mu$-distributed random matrix and $z$ is a $\nu$-distributed
random direction, then the direction 
\begin{equation}
{\mathscr A} (z) := \frac{a z + b}{c z + d}
\label{action}
\end{equation}
---of the vector obtained after $A$ has multiplied a vector of direction $z$---
is also $\nu$-distributed.
In the particular case where $\nu$ has a density, i.e.
$$
\nu (\d z) = f(z) \,\d z \,,
$$
it may be shown that
\begin{equation}
  f(z) = \int_{\text{SL} \left ( 2, {\mathbb R} \right )} \mu (\d A )\,
  \left ( f \circ {\mathscr A}^{-1} \right ) (z)\,
  \frac{\d {\mathscr A}^{-1}}{\d z} (z) \,.
\label{integralEquation}
\end{equation}
However, there is no systematic method for solving this integral equation. 

\subsection{The particular products of random matrices considered}
\label{mainResultsSubsection}
To describe them, let us first remark that every 
$A\in\text{SL}\left( 2, {\mathbb R} \right)$ has a unique Iwasawa
decomposition 
\begin{equation}
  A = 
  \begin{pmatrix}
    \cos \theta & -\sin \theta \\
    \sin \theta & \cos \theta
  \end{pmatrix}
  \begin{pmatrix}
    \e^{w} & 0 \\
    0 & \e^{-w}
  \end{pmatrix}
  \begin{pmatrix}
    1 & u \\
    0 & 1
  \end{pmatrix}
  \label{gramSchmidtDecomposition}
\end{equation}
for some $\theta$, $u,\, w \in {\mathbb R}$. This follows easily by
applying the familiar Gram--Schmidt algorithm to the columns 
of $A$. The three parameters in this decomposition have simple
geometrical meanings: $-\theta$ is the angle that the first column of
$A$ makes with the horizontal axis, $\e^{w}$ is its magnitude, and $u$
is related to the angle between the columns; in particular, $u=0$ if
and only if the columns are orthogonal. 

Now, suppose that these three parameters are independent random variables. 
We use the notation
$$
v \sim \text{\tt Exp}(r)
$$
to indicate that $v$ is a random variable with an exponential
distribution of parameter $r$, i.e. 
its density is given by 
$$
 r\, \e^{-r v }\, {\mathbf 1}_{(0,\infty)}(v)
  \:,
$$
where for every set $A\subset{\mathbb R}$, 
$$
  {\mathbf 1}_{A}(x) =
  \begin{cases}
    1 & \mbox{ for } x\in A \\
    0 & \mbox{ otherwise}
  \end{cases}
  \:.
$$
Also, $\delta_x$ will denote the discrete probability distribution on
${\mathbb R}$ with 
all the mass at $x$.
We shall provide an explicit formula for the $\mu$-invariant measure
of the product $\Pi_n$ when the matrices are independent draws from
the distribution $\mu$ of $A$ corresponding to either 
\begin{equation}
  \theta \sim \text{\tt Exp} (p)\,,\; 
  \pm u \sim \text{\tt Exp} (q)\,,\; 
  w \sim \delta_0\,.
  \label{frischLloydCase}
\end{equation}
or
\begin{equation}
  \theta \sim \text{\tt Exp} (p)\,, \; 
  u \sim \delta_0\,,\: 
  \pm w \sim \text{\tt Exp} (q)\,.
\label{susyCase}
\end{equation}

We shall also look at other closely related products: for instance,
products involving matrices of the form 
$$
A = \begin{pmatrix}
\cosh \theta & \sinh \theta \\
\sinh \theta & \cosh \theta
\end{pmatrix}
\begin{pmatrix}
\e^{w} & 0 \\
0 & \e^{-w}
\end{pmatrix}
\begin{pmatrix}
1 & u \\
0 & 1
\end{pmatrix}
$$
and we shall exhibit invariant measures for such cases too.

\subsection{The Schr\"{o}dinger equation with a random potential}
\label{schroedingerSubsection}
Our approach to computing the invariant measure will not make explicit
use of the integral equation (\ref{integralEquation}). Instead, we
shall exploit the fact that these products arise when solving the
Schr\"{o}dinger equation (in units such that
$\hbar=2m=1$)  
\begin{equation}
  -\psi''(x) + V(x) \,\psi (x) = E \psi(x)
\label{schroedingerEquation}
\end{equation}
for a given energy $E$ and a potential function $V$ that vanishes
everywhere except on a countable set of points $\{x_j\}$. Physically 
speaking, one can think of $\psi$ as the wave function of a quantum
particle in a crystal with impurities; 
the effect of the impurity located at $x_j$ is modelled by
the boundary condition
$$
\begin{pmatrix}
\psi'(x_j+) \\
\psi (x_j+)
\end{pmatrix}
= B_j 
\begin{pmatrix}
\psi'(x_j-) \\
\psi (x_j-)
\end{pmatrix}
$$
where $B_j \in \text{SL}(2,{\mathbb R})$. The potential $V$ is
therefore a sum of simpler 
potentials, one for each pair $(x_j,B_j)$, known variously as {\em
  point scatterers}, {\em generalised contact scatterers} or {\em
  pointlike scatterers} \cite{ADK,AGHH,CheShi04,CheHug93,Ex,Se}. 
The case (\ref{frischLloydCase}) corresponds to the 
disordered version of the familiar Kronig--Penney model \cite{KP} considered
by Frisch \& Lloyd \cite{FL} and Kotani \cite{Ko}. 
The case (\ref{susyCase}) corresponds to a ``supersymmetric version'' of
the same model, in which the Schr\"{o}dinger operator factorises as
\begin{multline}
  -\frac{\d^2}{\d x^2} + V(x) 
  = -\frac{\d^2}{\d x^2} + W(x)^2 - W'(x)
  \\
  = \left [-\frac{\d}{\d x} + W(x) \right ] 
    \left [ \frac{\d}{\d x} + W(x) \right ] 
\label{eq:Hsusy}
\end{multline}
and the {\em superpotential} $W$ is of the Kronig--Penney type.
Such a supersymmetric
Hamiltonian is related to the square of a Dirac 
operator with a random mass $W$--- a model that is of independent
interest in many 
contexts of condensed matter physics
\cite{BouComGeoLeD90,ComTex98,CKS,GurCha03,TexHag09}.

The strategy for calculating $\nu$ is based on the observation that it is also the stationary distribution of a certain Markov
process $\{ z(x) \}$, where
$$
z := \frac{\psi'}{\psi}
$$ 
is the Riccati variable associated with the Schr\"{o}dinger
equation. 
In the particular case where 
$$
x_{j+1} - x_j \sim \text{Exp}(p)
$$
and the $B_j$ are independent and identically distributed random variables in $\text{SL} (2,{\mathbb R})$,
one can, following Frisch \& Lloyd \cite{FL}, show that the density of the stationary distribution satisfies
a certain integro-differential equation. 
The cases 
(\ref{frischLloydCase}) and (\ref{susyCase}) share a special feature:
the distribution of the $B_j$ is such that 
the integro-differential equation may be reduced to  a {\em differential}
equation. Furthermore, this differential equation is simple enough to
admit an exact solution in terms of elementary functions. 
 
The idea of using the Riccati variable to study disordered systems
goes back to Frisch \& Lloyd \cite{FL}. The well-known ``phase
formalism'' introduced in~\cite{AntPasSly81,LGP} is another version of
the same idea.
The trick that allows one to express the equation for the stationary
distribution of the Riccati variable
in a purely differential form is borrowed from Nieuwenhuizen's work
\cite{Ni} on the particular 
case (\ref{frischLloydCase}), in which the Dyson--Schmidt method is
used to compute the Lyapunov exponent directly from a so-called
characteristic function. The same trick has been used by others in
various contexts \cite{CPY,GP,MTW}. The key fact is that
the density of the exponential distribution satisfies a linear differential
equation with constant coefficients. Our results on products of
matrices therefore admit a number of extensions; for instance when
$\pm v$ (or $\pm w$) has, say, a gamma or a Laplace (i.e. piecewise
exponential) distribution.
%
%
One difficulty that arises with these distributions is that the 
differential equation for the invariant density is then of second
or higher order. This makes it harder to identify 
the relevant solution; furthermore, this solution is 
seldom expressible in terms of elementary functions. Without aiming at
an exhaustive treatment, we shall have occasion to illustrate some 
of these technical difficulties.

\subsection{Outline of the paper}
\label{outlineSubsection}
The remainder of the paper is as follows:  in \S \ref{pointSection},
we review the concept of point scatterer. The 
Frisch--Lloyd equation for the stationary density of the Riccati variable is
derived in \S \ref{generalisedKronigPenneySection}. In 
\S \ref{invariantSection}, we study particular choices of random point 
scatterers for which the Frisch--Lloyd equation can be reduced to a
purely differential form.  We can solve this equation in some cases 
and these results are then translated 
in terms of invariant measures for products of random
matrices. Some possible extensions of our results are
discussed in~\S\ref{extensionSection}. We end the paper with a few
concluding remarks in~\S\ref{sec:Conclusion}.

\section{Point scatterers}
\label{pointSection}
Let $u \in {\mathbb R}$ and let $\delta$ denote the Dirac delta. The
Schr\"{o}dinger equation with the potential 
$$
V(x) = u \,\delta(x)
$$
can be expressed in the equivalent form
\begin{equation}
  -\psi'' = E \psi\,, \quad x \ne 0\,,
  \label{schroedingerEquationWithPointInteraction}
\end{equation}
and
\begin{equation}
  \psi(0+) = \psi(0-)\,, \quad \psi'(0+) = \psi'(0-) + u\, \psi(0-)\,.
  \label{deltaInteraction}
\end{equation}
This familiar ``delta scatterer'' is a convenient idealisation for a
short-range, highly localised potential.  

A (mathematically) natural generalisation of this scatterer is
obtained when the boundary condition (\ref{deltaInteraction}) is
replaced by
\begin{equation}
  \begin{pmatrix}
    \psi'(0+) \\
    \psi (0+)
    \end{pmatrix}
    = B 
    \begin{pmatrix}
    \psi'(0-) \\
    \psi (0-)
  \end{pmatrix}
  \label{pointInteraction}
\end{equation}
where $B$ is some $2\times2$ matrix. We shall refer to $B$  
as the ``boundary matrix''.
In order to ascertain what boundary matrices yield a Schr\"{o}dinger operator
with a self-adjoint extension, 
we start with the observation that the probability current associated with
the wavefunction is proportional to
$$
  \begin{pmatrix}
    \, \overline{\psi'(x)} & \overline{\psi(x)} \,
  \end{pmatrix}
  \begin{pmatrix}
    0 & -1 \\
    1 & 0
  \end{pmatrix}
    \begin{pmatrix}
    \psi'(x)  \\
    \psi (x) 
  \end{pmatrix}
$$
where the bar denotes complex conjugation. 
The requirement that the probability current should be the same on both sides of the 
scatterer translates into the following condition on $B$ \cite{CheShi04}:
$$
  B^\dagger   
  \begin{pmatrix}
    0 & -1 \\
    1 & 0
  \end{pmatrix}
  B
  =
  \begin{pmatrix}
    0 & -1 \\
    1 & 0
  \end{pmatrix}
$$
where the dagger denotes hermitian transposition. Equivalently,
$$
b_{11}\overline{b_{22}}-b_{21}\overline{b_{12}}=1\;\;\text{and}\;\;
\text{Im}(b_{11}\overline{b_{21}})=\text{Im}(b_{22}\overline{b_{12}})=0\,.
$$
It is easily seen that this forces 
\cite{ADK,Se}
$$
  \e^{-\i \chi} B \in \text{SL} \left (2, {\mathbb R} \right )
$$
for some real number $\chi$. As discussed in
Appendix \ref{app:ScatteringTransfer}, for the purposes of
this paper there is no loss of generality in setting $\chi = 0$
and restricting our 
attention to the case of {\em real} boundary matrices.

We write
\begin{equation}
  V (x) = \sigma_B (x)
  \label{pointInteractionPotential}
\end{equation}
for the potential with these properties, and call it a {\em point scatterer} (at the origin) or, as it is also known, a generalised
contact scatterer or pointlike scatterer \cite{ADK,AGHH,Ex,Se}. 
We remark that the Riccati variable $z = \psi'/\psi$ of the Schr\"{o}dinger equation with this potential satisfies
\begin{equation}
  z' = - \left ( E+z^2 \right )\,, \quad x \ne 0\,,
  \label{riccatiEquationWithPointInteraction}
\end{equation}
and
\begin{equation}
  z ( 0+) = {\mathscr B} \left ( z(0-) \right )
  \label{riccatiBoundaryConditionAtAnInteraction}
\end{equation}
where ${\mathscr B}$ is the linear fractional transformation
associated with the matrix $B$: 
\begin{equation}
  {\mathscr B}(z) = \frac{b_{11} z + b_{12}}{b_{21} z + b_{22}}\,.
  \label{linearFractionalTransformation}
\end{equation}
The fact that $B\in\text{SL}(2,{\mathbb R})$ ensures that ${\mathscr B}$
is invertible.

In order to gain some insight into the possible physical  significance of the
boundary matrix $B$, we
set $E=k^2$, $k > 0$, and look for solutions of Equations
(\ref{schroedingerEquationWithPointInteraction}) and
(\ref{pointInteraction}) of the form 
\begin{equation}
  \psi(x) =
  \begin{cases}
   a^\text{in}_- \, \e^{\i kx} + a^\text{out}_- \, \e^{-\i kx} &
    \text{for $x<0$} \\
   a^\text{out}_+  \, \e^{\i kx} + a^\text{in}_+ \, \e^{-\i kx} &
    \text{for $x>0$}
  \end{cases}\,.
  \label{eq:ScatteringState}
\end{equation}
By definition, the scattering matrix $S$ relates the incoming
amplitudes to the outgoing amplitudes via
\begin{equation}
  \begin{pmatrix}
  a^\text{out}_- \\
  a^\text{out}_+
  \end{pmatrix}
  = S \begin{pmatrix}
  a^\text{in}_- \\
  a^\text{in}_+
  \end{pmatrix}\,.
  \label{eq:ScatteringMatrix}
\end{equation}
Hence
\begin{multline}
  \notag
  S = \frac{1}{b_{21} k^2 + \i k (b_{11}+b_{22}) - b_{12}} \\
  \times \begin{pmatrix} 
  b_{21} k^2 - \i k (b_{22}-b_{11})+b_{12} & 2 \i k\\
  2 \i k ( b_{11} b_{22}-b_{12} b_{21}) & b_{21} k^2 + \i k (b_{22}-b_{11})+b_{12} 
  \end{pmatrix}\,.
\end{multline}
The relationship between boundary and scattering matrices is discussed at greater length in 
Appendix \ref{app:ScatteringTransfer}.

\begin{example}
For the delta scatterer defined by (\ref{deltaInteraction}),
$$
B = \begin{pmatrix}
1 & u \\
0 & 1
\end{pmatrix}\,.
$$ 
The wave function is continuous at the origin, but its derivative
experiences a jump proportional to the value of the wave function
there. 
We have
$$
  S = \frac{1}{2 \i k -u} \begin{pmatrix}
  u & 2 \i k \\
  2 \i k & u 
  \end{pmatrix}
  \quad \text{and} \quad
  {\mathscr B}(z) = z+u\,.
$$
The fact that
$$
\lim_{u\rightarrow\pm\infty}S=-I\,,
$$
where $I$ is the
identity matrix, indicates that the limiting case
of an infinitely large ``impurity strength'' $u$
corresponds to imposing a Dirichlet
boundary condition at the scatterer's position.
\label{deltaExample}
\end{example}

\begin{example}
The ``delta--prime'' scatterer (see for instance \cite{AGHH,Se}) is defined
by
$$
B = \begin{pmatrix}
1 & 0\\
v & 1
\end{pmatrix}
$$ 
where $v\in{\mathbb R}$. Now it is the derivative of the wave
function that is continuous at the origin, and the wave function that  
jumps: 
$$
\psi(0+)-\psi(0-)=v\,\psi'(0)\,.
$$ 
We emphasise that, in spite of its (widely used) name, 
the delta-prime scatterer {\em does not} 
correspond to using the
distributional derivative $\delta'$ as a potential \cite{ADK}.

We have
$$
  S = \frac{1}{2 \i + v k} \begin{pmatrix}
  v k & 2 \i \\
  2 \i & v k 
  \end{pmatrix}
  \quad \text{and} \quad
  {\mathscr B}(z) = \frac{z}{v z+ 1}\,.
$$
The fact that 
$$
\lim_{v\rightarrow\pm\infty}S=+I
$$ 
indicates that a Neumann
boundary condition is obtained in the limit of infinite strength $v$.

The question of the possible physical significance of the delta-prime scatterer
was considered by Cheon \& Shigehara \cite{CS}, who 
showed that it can in principle be
``realised'' by taking an appropriate limit of three neighbouring 
delta scatterers.
\label{deltaPrimeExample}
\end{example}

\begin{example}
Let $w \in {\mathbb R}$ and 
\begin{equation}
  B = \begin{pmatrix}
  \e^{w} & 0\\
  0 & \e^{-w}
  \end{pmatrix}\,.
  \label{eq:matrixBsusy}
\end{equation}
In this case, the scatterer produces a discontinuity in both the wave function and its derivative.
As pointed out in \cite{CheHug93},
the Schr\"{o}dinger
equation (\ref{schroedingerEquationWithPointInteraction}) can be
recast as the first-order 
system
\begin{align*}
- \psi' - W \psi &= k \phi \\
\phi' -  W \phi &= k \psi
\end{align*}
with 
$$
W(x) = w \,\delta(x)\,.
$$
The meaning of these equations becomes clear if we introduce an
integrating factor: 
\begin{align*}
- \frac{\d}{\d x} \left [ \exp \left (  \int_{-\infty}^x W(y)\,\d y \right ) \psi \right ] &= k \exp \left ( \int_{-\infty}^x W(y)\,\d y \right ) \phi \\
\frac{\d}{\d x} \left [ \exp \left (  -\int_{-\infty}^x W(y)\,\d y \right ) \phi \right ] &= k \exp \left ( -\int_{-\infty}^x W(y)\,\d y \right ) \psi\,.
\end{align*}
We call this scatterer the {\em supersymmetric scatterer}.
We have
$$
  S = \begin{pmatrix}
  \tanh w & \text{sech} \,w \\
  \text{sech} \,w & -\tanh w 
  \end{pmatrix}
  \quad \text{and} \quad
  {\mathscr B}(z) = \e^{2 w} z\,.
$$
Hence the scattering is independent of the wave number $k$--- a property consistent with the observation, made in
Albeverio {\em et al}\,\cite{ADK}, that diagonal matrices (are the only matrices in $\text{SL}(2,{\mathbb R})$ that)
yield boundary conditions invariant
under the scaling
$$
\psi (x) \mapsto \sqrt{\lambda} \,\psi (\lambda x)\,, \;\; \lambda > 0\,.
$$
However, in
contrast with the previous examples, the scattering is asymmetric, i.e. not 
invariant under the transformation $x\mapsto-x$.  
The limit of infinite strength $w$ has a clear
interpretation: it corresponds to a Neumann boundary condition on the left
of the barrier, and to a Dirichlet condition on the right.
\label{supersymmetricExample}
\end{example}

\begin{example}
Let
$$
B = \begin{pmatrix}
\e^{w} & 0 \\
0 & \e^{-w}
\end{pmatrix}
\begin{pmatrix}
1 & u \\
0 & 1
\end{pmatrix}
\,.
$$
This point scatterer can be thought of as two neighbouring
scatterers--- a supersymmetric scatterer of strength $w$ 
on the right, and a delta scatterer of strength $u$
on the left--- in the limit as the distance $\varepsilon$ separating
them tends to $0$; see Figure \ref{doubleImpurityFigure}. For want of
a better name, 
we shall refer to it as the {\em double impurity}.
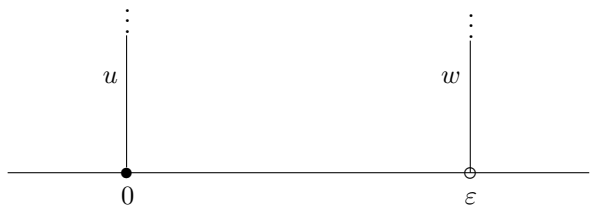
\begin{figure}[htbp]
\vspace{3cm} 
\begin{picture}(0,0) 
\put(-100,20){\line(1,0){220}}
\put(-55,20){\circle*{4}}
\put(-57,8){$0$}
\put(-64,54){$u$}
\put(75,20){\circle{4}}
\put(73,8){$\varepsilon$}
\put(64,54){$w$}
\put(-55,22){\line(0,1){50}}
\put(-56,73){\vdots}
\put(75,20){\line(0,1){50}}
\put(74,71){\vdots}
\end{picture}  
\caption{The double impurity: 
the empty dot corresponds to the location of
a supersymmetric scatterer of strength $w$ while the black dot  
corresponds to the location of a delta scatterer of strength~$u$.} 
\label{doubleImpurityFigure} 
\end{figure}
\label{doubleImpurityExample}
\end{example}
We have
$$
S = \frac{1}{2 \i k \cosh w - u \e^{w}} \begin{pmatrix}
2 \i k \sinh w + u \e^{w} & 2 \i k \\
2 \i k & -2 \i k \sinh w + u \e^{w} 
\end{pmatrix}
$$
and
$$
{\mathscr B} (z) = \e^{2 w} (z+u)\,.
$$

This particular scatterer is interesting for the following reason: the Iwasawa
decomposition (\ref{gramSchmidtDecomposition}) implies that
{\em any} point scatterer for a real boundary matrix can be
thought of  as a double impurity ``up to a rotation''.
For 
example, the boundary matrix for the delta-prime scatterer may be
decomposed as
$$
\begin{pmatrix}
1 & 0 \\
v & 1
\end{pmatrix}
= \begin{pmatrix}
\cos \theta & -\sin \theta \\
\sin \theta & \cos \theta
\end{pmatrix}
\begin{pmatrix}
\e^{w} & 0 \\
0 & \e^{-w}
\end{pmatrix}
\begin{pmatrix}
1 & u \\
0 & 1
\end{pmatrix}
$$
with
$$
\theta= \arctan v, \;\; w=\frac{1}{2} \ln(1+v^2)\;\; \text{and}\;\;
u=\frac{v}{1+v^2}\,. 
$$
We shall return to this point in the next section.

\section{A generalised Kronig--Penney model with disorder}
\label{generalisedKronigPenneySection}
In this section, we elaborate the correspondence between disordered systems
with point scatterers and products of random matrices. Then, for a
particular type of disorder, 
we show how, following Frisch \& Lloyd \cite{FL}, one can derive a
useful equation for the 
stationary density of the Riccati variable associated with the system.

\subsection{The generalised Kronig--Penney model}
\label{generalisedSubsection}
Given a sequence 
$$
\{ B_j \} \subset \text{SL}(2,{\mathbb R})
$$ 
and an
increasing sequence 
$\{ x_j \}$ of non-negative numbers, we call the equation
(\ref{schroedingerEquation}) with the potential
\begin{equation}
  V(x) = \sum_{j=1}^\infty  \sigma_{B_j} (x-x_j)
  \label{generalisedKronigPenneyPotential}
\end{equation}
a {\em generalised Kronig--Penney model}. 
The notation $\sigma_B(x)$ was defined in
Equation~\eqref{pointInteractionPotential}. 

Let us consider first the case where the energy is positive, i.e. $E = k^2$, $k >0$. In principle, one could dispense
with the parameter $k$ and set it to unity by rescaling $x$ but, as we shall see later in \S \ref{halperinSubsection}, there is some advantage
in making the dependence on the energy explicit. For $x_j < x < x_{j+1}$, the
solution is given by
\begin{multline}
  \begin{pmatrix}
    \psi' (x) \\
    \psi (x)
  \end{pmatrix}
  = 
  \begin{pmatrix}
     \sqrt{k} & 0 \\
     0 & \frac{1}{\sqrt{k}}
  \end{pmatrix}
  \begin{pmatrix}
     \cos \left ( k [x-x_j] \right ) & -\sin \left ( k [x-x_j] \right ) \\
     \sin \left ( k [x-x_j] \right ) & \cos \left ( k [x-x_j] \right )
  \end{pmatrix} \\
  \notag \times
  \begin{pmatrix}
    \frac{1}{\sqrt{k}} & 0 \\
     0 & \sqrt{k}
  \end{pmatrix} B_j 
  \begin{pmatrix}
     \psi'( x_j- ) \\
     \psi (x_j- )
  \end{pmatrix}
  \label{solutionOftheGeneralisedKronigPenneyModel}\,.
\end{multline}
By recurrence, we obtain the solution for every $x > 0$ in terms
of a product of matrices. In particular,
\begin{equation}
\begin{pmatrix}
\psi' (x_{n+1}-) \\
\psi (x_{n+1}-)
\end{pmatrix}
= A_{n} A_{n-1} \cdots A_{1} 
\begin{pmatrix}
\psi' (x_1-) \\
\psi (x_1-)
\end{pmatrix}
\label{generalisedKronigPenneySolution}
\end{equation}
where
\begin{equation}
A_j = \begin{pmatrix}
\sqrt{k} & 0 \\
0 & \frac{1}{\sqrt{k}}
\end{pmatrix}
\begin{pmatrix}
\cos \left ( k \theta_j \right ) & -\sin \left ( k \theta_j \right ) \\
\sin \left ( k \theta_j \right ) & \cos \left ( k \theta_j \right )
\end{pmatrix} 
\begin{pmatrix}
\frac{1}{\sqrt{k}} & 0 \\
0 & \sqrt{k}
\end{pmatrix} B_j 
\label{generalisedKronigPenneyMatrixForPositiveEnergy}
\end{equation}
and
$$
\theta_j := x_{j+1}-x_j\,.
$$
Thus, for instance, we see that a product of matrices of the form
(\ref{gramSchmidtDecomposition}) 
corresponds to a generalised Kronig--Penney model of unit energy in
which the $\sigma_{B_j}$ are double impurities.
It is worth emphasising this point: between impurities, the
Schr\"{o}dinger operator 
itself produces the ``rotation part'' of the matrices in the
product. Therefore, in order to associate 
a quantum model to the most general product of matrices, it
is sufficient to use a potential made up of (suitably
spaced) double impurities.

The case of negative energy, i.e. $E =-k^2$, $k > 0$, is also of
mathematical interest.  Then
Equation (\ref{generalisedKronigPenneySolution}) holds with
\begin{equation}
  A_j = \begin{pmatrix}
  \sqrt{k} & 0 \\
  0 & \frac{1}{\sqrt{k}}
  \end{pmatrix}
  \begin{pmatrix}
  \cosh \left ( k \theta_j \right ) & \sinh \left ( k \theta_j \right ) \\
  \sinh \left ( k \theta_j \right ) & \cosh \left ( k \theta_j \right )
  \end{pmatrix} 
  \begin{pmatrix}
  \frac{1}{\sqrt{k}} & 0 \\
  0 & \sqrt{k}
  \end{pmatrix}B_j \,.
  \label{generalisedKronigPenneyMatrixForNegativeEnergy}
\end{equation}

\subsection{The generalised Frisch--Lloyd equation}
\label{frischLloydEquationSubsection}
It is physically reasonable to assume that the scatterers are randomly,
independently and uniformly distributed. We denote by $p$ the mean
density of impurities. If we label the scatterers in order of increasing
position along the positive semi-axis, so that $x_j$ denotes the position of the $j$th impurity, then
$$
0 < x_1<x_2<x_3<\cdots
$$
and the spacings 
between consecutive
scatterers are independent and have the same
exponential distribution, i.e. 
\begin{equation}
  \theta_j \sim \text{\tt Exp}(p)\,, \quad p > 0\,.
  \label{exponentiallyDistributedTheta}
\end{equation}
For this distribution of the $\theta_j$,
$$
n(x) := \# \left \{ x_j :\; x_j < x \right \}
$$
is the familiar Poisson process.  

We
shall be interested in the statistical behaviour of the Riccati variable
\begin{equation}
  z(x) = \frac{\psi'(x)}{\psi(x)}\,.
  \label{definitionRiccatiEquation}
\end{equation}
Its evolution is governed by
\begin{equation}
  z' = - \left ( z^2 + E \right )\,, \quad x \notin \{ x_j \}\,,
  \label{freeRiccatiEquation}
\end{equation}
and 
\begin{equation}
  z(x_j+) = {\mathscr B}_j \left ( z(x_j-) \right )\,, 
  \quad j \in {\mathbb N}\,.
  \label{riccatiJump}
\end{equation}
The ``lack of memory'' property of the exponential distribution
\eqref{exponentiallyDistributedTheta} implies that
the process $\{ z(x) \}$ thus defined is Markov.

It should be clear from \S \ref{productSubsection} and the
previous subsection that, if we set $k=1$, then the invariant measure
$\nu$ associated with the product
(\ref{generalisedKronigPenneySolution}) 
is precisely the stationary distribution of the Riccati variable. So
we shall look for particular cases where this stationary distribution
may be obtained in analytical form. 

To simplify matters, we also suppose in the first instance that the
$B_j$ are all the same, deterministic, and we drop the subscript.

Let $f(z;x)$ be the density of the distribution
of the Riccati variable.
Let $h>0$ and let $d z$ denote an interval of infinitesimal length $\d z$ centered on the number
$z$. Then
\begin{multline}
  \notag
  f(z;x+h) \,\d z = {\mathbb P} \left ( z(x+h) \in dz \right ) \\
  = \sum_{\ell=0}^\infty 
    {\mathbb P} \left ( z(x+h) \in dz \,\Bigl | \,n(x+h)-n(x) = \ell
    \right ) 
   {\mathbb P} \left ( n(x+h)-n(x) = \ell \right ) \,.
\end{multline}
It is well-known (see \cite{Fe}) that, with an error of order $o(h)$
as $h \rightarrow 0+$, 
\begin{equation}
\notag
{\mathbb P} \left ( n(x+h) - n(x) = \ell \right ) = \begin{cases}
1-p \,h & \text{if $\ell=0$} \\
p \,h & \text{if $\ell = 1$} \\
0 & \text{if $\ell > 1$}
\end{cases}
\end{equation}
and so
\begin{multline}
  f(z;x+h) \,\d z = 
  {\mathbb P} \left ( z(x+h) \in dz \,\Bigl | \,n(x+h)-n(x) = 0 \right) 
  (1- p h) \\
  + {\mathbb P} \left ( z(x+h) \in dz \,\Bigl | \,n(x+h)-n(x) = 1 \right )
  p h + \d z \,o (h) \quad \text{as $h \rightarrow 0+$}\,.
\label{sumOfProbabilities}
\end{multline}
The condition $n(x+h)-n(x)=0$ means that no $x_j$ lies in $(x,x+h)$,
and so implies that the Riccati variable is governed solely by 
the differential equation (\ref{freeRiccatiEquation}) in this
interval. Therefore, 
the first conditional probability on the right-hand side of
(\ref{sumOfProbabilities}) equals 
$$
  f \left ( z+[z^2+E]h ; x \right ) 
  \left [ 1 + 2 z h + o(h) \right ] \d z \quad \text{as $h \rightarrow 0+$}\,.
$$
The condition $n(x+h)-n(x)=1$ means that exactly one of the $x_j$ lies in $(x,x+h)$, and so the Riccati variable experiences a jump
defined by Equation (\ref{riccatiJump}) in this interval. A simple calculation then yields, for the second conditional
probability on the right-hand side of (\ref{sumOfProbabilities}),
\begin{multline}
  \notag
  {\mathbb P} \left ( z(x+h) \in dz \,\Bigl | \,n(x+h)-n(x) = 1 \right) 
  = {\mathbb P} \left ( {\mathscr B} \left (z(x_j-) \right ) \in dz \right )
  + \d z \,O(h) \\
  = f \left( {\mathscr B}^{-1}(z) ; x \right) 
  \frac{\d {\mathscr B}^{-1}}{\d z} (z) \,\d z \,\left [ 1 + O(h)
  \right ] 
  \quad \text{as $h \rightarrow 0+$}\,.
\end{multline}
After reporting these results in Equation (\ref{sumOfProbabilities})
and taking the limit as $h \rightarrow 0+$, we obtain a generalisation
of the equation (6.69) in \cite{LGP}, \S 6.7:
\begin{equation}
  \notag
  \frac{\partial f}{\partial x} (z;x) = 
  \frac{\partial}{\partial z} \left [ (z^2+E) \,f(z;x) \right ] + p 
  \left [ f \left ( {\mathscr B}^{-1}(z) ; x \right ) 
  \frac{\d {\mathscr B}^{-1}}{\d z} (z) - f(z;x) \right ]\,.
\end{equation}
The stationary distribution, denoted again $f=f(z)$, therefore satisfies
$$
\frac{\d}{\d z} \left [ (z^2+E) \,f(z) \right ] + p 
\left [ f \left ( {\mathscr B}^{-1}(z) \right ) \frac{\d {\mathscr B}^{-1}}{\d z} (z) - f(z) \right ]= 0\,.
$$

More generally, if we permit the $B_j$ to be independent random variables with a common distribution denoted $\kappa$, then it is straightforward
to derive the equation 
\begin{equation}
  \frac{\d}{\d z} \left [ (z^2 + E) \,f(z) \right ] 
  + p \int_{\text{SL}(2, \mathbb R)} \kappa ( \d B)\, 
  \left [ f \left ( {\mathscr B}^{-1} (z) \right )
   \frac{\d {\mathscr B}^{-1}}{\d z}(z) - f(z) \right ]=0\,.
  \label{frischLloydEquationInDifferentialForm}
\end{equation}
By integrating with respect to $z$, we obtain
\begin{equation}
  (z^2 + E)\,f(z) + p \int_{\text{SL}(2, \mathbb R)}  \kappa ( \d B) 
  \int_{z}^{{\mathscr B}^{-1} (z)}\d t\,  f(t) =N\,.
  \label{frischLloydEquation}
\end{equation}
The constant of integration $N$ in this equation depends on $E$; as will be explained shortly,
it represents the {\em integrated density of states}
per unit length of the Schr\"odinger Hamiltonian for the
potential \eqref{generalisedKronigPenneyPotential}
\cite{FL,Ko,LGP}. 

We shall refer to Equation
(\ref{frischLloydEquationInDifferentialForm}), or to its integrated
version 
(\ref{frischLloydEquation}),
as the (generalised) {\em Frisch--Lloyd equation}. 
In the following sections, we shall consider again the particular point scatterers described in \S \ref{pointSection}, and 
exhibit choices of the measure $\kappa$ for which
this equation can be converted
to a differential equation.  

\subsection{The qualitative behaviour of the Riccati variable}
\label{qualitativeSubsection}
\begin{figure}[htbp]
\includegraphics[scale=1]{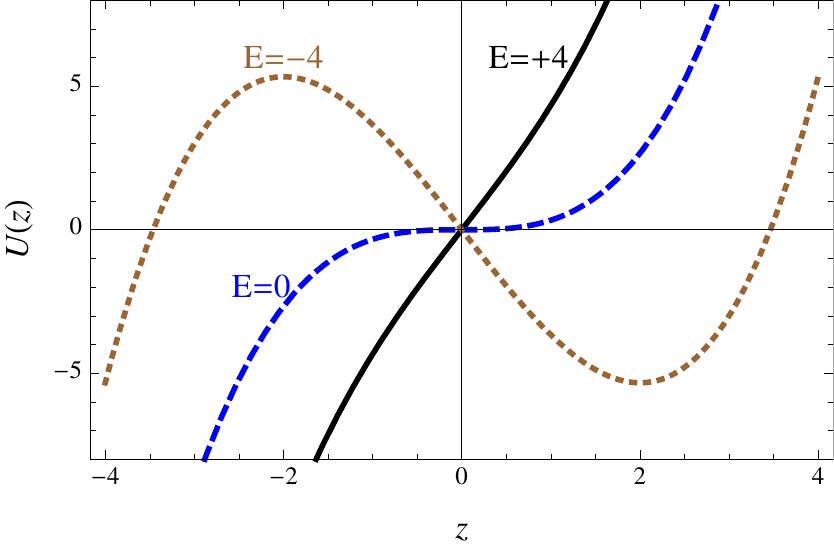}
\caption{The ``potential'' $U(z)$ associated with the unperturbed
  Riccati equation 
$z' = -U'(z) = - (z^2+E)$.} 
\label{fig:pot_U} 
\end{figure}

It is instructive to think of the Riccati equation (\ref{freeRiccatiEquation}) in the absence of scatterers
as an autonomous system describing the motion of
a fictitious ``particle'' constrained to roll along the ``potential'' curve
$$
U(z) =  E z + \frac{z^3}{3}
$$
in such a way that its ``velocity'' at ``time'' $x$ and ``position'' $z$ is given by the slope $-U'(z)$; see
Figure \ref{fig:pot_U}.
We may regard the occurence of the jumps
in Equation (\ref{riccatiJump}) as a perturbation of this autonomous system, and the intensity $p$ of the Poisson process as the perturbation parameter. 

Let us consider first the unperturbed system (i.e. $p=0$).
For $E>0$, the system has no equilibrium point: the particle rolls
down to $-\infty$, and re-appears immediately at $+\infty$, reflecting
the fact that the solution $\psi$ of the corresponding Schr\"{o}dinger
equation has a zero at the ``time'' $x$ when the particle escapes to infinity. 
This behaviour of the Riccati variable indicates that every $E>0$
belongs to the spectrum of the Schr\"{o}dinger operator.

Equation \eqref{freeRiccatiEquation} gives the ``velocity'' of the
fictitious particle as a function of its position. Hence  the ``time'' taken to go from $+\infty$ to
$-\infty$ is
$$
-\int_{+\infty}^{-\infty}\frac{\d{z}}{z^2+k^2}=\frac{\pi}{k}\,.
$$
On the other hand, 
the solution of the Frisch-Lloyd equation for $E>0$ and $p=0$ is the Cauchy
density
\begin{equation}
f(z) = \frac{N}{z^2+k^2} \;\;\text{with}\;\; N = \frac{k}{\pi}\,.
\label{cauchyDensity}
\end{equation}
Therefore the normalisation constant
$N$ may be interpreted 
as the reciprocal of the ``time'' that the particle takes to run through ${\mathbb R}$.
Another equivalent interpretation is as follows:
recall that, when the particle escapes
to $-\infty$, it is immediately re-injected at $+\infty$ to commence 
a new journey through ${\mathbb R}$. 
$N$ may therefore also be viewed as  
the {\em current} of the fictitious particle \cite{LGP}, and
the Rice formula 
$$
\lim_{z\to\pm\infty}z^2f(z)=N
$$ 
can be understood as expressing a relation between the stationary
distribution and a current of probability.
This current equals the number
of infinitudes of $z(x)$--- i.e. the number
of nodes of the wavefunction $\psi(x)$---  per unit length. By the familiar oscillation
theorem of Sturm--Liouville theory, 
it is therefore the same as the integrated density of
states per unit length of the corresponding Schr\"odinger Hamiltonian.

By contrast, in the case $E=-k^2 < 0$, $k>0$, the unperturbed system
has an unstable equilibrium point at $-k$, and a stable equilibrium
point at $k$. 
Unless the particle starts from a position on 
the left of
the unstable equilibrium, 
it must tend asymptotically to the stable
equilibrium point. The fact that the particle cannot reach infinity
more than once indicates that  
the spectrum lies entirely in ${\mathbb R}_+$. The solution of the
Frisch--Lloyd equation is 
$$
  f(z) = \delta(z-k)\,.
$$ 

Let us now consider how the occurence of jumps can affect the
system. For $E>0$, the jumps
defined by Equation (\ref{riccatiJump}), as long as they
are finite and  
infrequent (i.e. $p$ 
is small), cannot prevent the particle from visiting $-\infty$
repeatedly; the system should therefore behave in much the same way as
in the unperturbed case, 
and we expect the density $f$ to be Cauchy-like.
In particular, the interpretation of the normalisation constant in terms of a probability
current remains valid 
because, for $z$ large enough, the deterministic part
\eqref{freeRiccatiEquation} of the evolution of the Riccati variable
dominates the stochastic part (\ref{riccatiJump}). 
The situation for $E<0$ is more complicated.
Roughly speaking,
{\em positive} jumps, i.e. discontinuous increases of $z$, enable
the particle to make excursions to the right of the stable equilibrium
point $z=k$, but the particle can never overcome the 
infinite barrier and so it rolls back down towards $k$. On the other hand,
{\em negative} jumps, i.e. discontinuous decreases of $z$, enable
the particle to make excursions to the left of $k$. If the jump is
large enough, the particle can overcome the potential barrier at $-k$
and escape to 
$-\infty$, raising the possibility that part of the spectrum of the
Schr\"{o}dinger operator lies in ${\mathbb R}_-$. For small $p$, we
expect 
the density $f$ to be large in the neighbourhood of $z=k$.

We shall return to this useful particle analogy in later sections when
we examine the detailed behaviour of the Riccati variable for specific 
random point scatterers. 

\subsection{The reduced Lyapunov exponent}
\label{lyapunovSubsection}
Knowing the density of the invariant measure, the calculation of the
Lyapunov exponent reduces, thanks to Formula
(\ref{furstenbergFormula}), to the 
evaluation of a multiple integral. We will show in this subsection
that, if $A$ is of the form
(\ref{generalisedKronigPenneyMatrixForPositiveEnergy}) and the
boundary matrix $B$ is {\em triangular}, then this formula may be
greatly simplified.  As pointed out in \S \ref{generalisedSubsection}, 
there is no loss of generality in setting $E=k^2=1$ since the
parameter $k$ may be re-introduced subsequently by rescaling.
This has the advantage of making the calculation simpler.

For definiteness, let us begin with the upper triangular case,
i.e. $b_{21} = 0$. Then the density of the invariant measure satisfies
the Frisch--Lloyd equation
(\ref{frischLloydEquationInDifferentialForm}), again with $E=1$, and we have
\begin{multline}
  \notag
  \left | A \begin{pmatrix}
  z \\ 1
  \end{pmatrix}
  \right |^2 
  = \left | B \begin{pmatrix}
  z \\ 1
  \end{pmatrix}
  \right |^2  = 
  \left ( b_{11} z + b_{12} \right )^2 + \left ( b_{21} z + b_{22} \right )^2 \\
  = \left( b_{21} z + b_{22} \right)^2 \left[ 1 + {\mathscr B}(z)^2 \right] 
  = b_{22}^2 \left [ 1 + {\mathscr B}(z)^2 \right]\,.
\end{multline}
Hence
\begin{multline}
\notag
   \gamma_\mu 
  = \frac{1}{2} \int_{\mathbb R} \d z \int_{\text{SL} (2,\mathbb R )}\kappa (\d B)\,
  \ln \frac{\left | A \begin{pmatrix}
  z \\ 1
  \end{pmatrix}
  \right |^2}{\left | \begin{pmatrix}
  z \\ 1
  \end{pmatrix} \right |^2} \,f(z) \\ 
  = \frac{1}{2} \int_{\mathbb R} \d z \int_{\text{SL} (2,\mathbb R )}\kappa (\d B)\, 
\ln \frac{b_{22}^2 \left [ 1 + {\mathscr B}(z)^2 \right ]}{1+z^2}  \,f(z) 
\end{multline}
and, after some re-arrangement,
\begin{multline}
 \gamma_\mu  = {\mathbb E} \left ( \ln | b_{22} | \right ) + \frac{1}{2} \int_{\mathbb R} \d z \int_{\text{SL}(2,\mathbb R)}\kappa (\d B)\, \ln \left [ 1 + {\mathscr B}(z)^2 \right ] \,f(z) \\
- \frac{1}{2} \int_{\mathbb R} \d z \int_{\text{SL}(2,\mathbb R)}\kappa (\d B)\, \ln ( 1+z^2 )\,f(z)\,.
\label{lyapunovEquality}
\end{multline}
Consider the second term on the right-hand side of the last equality:
by changing the order of integration, and making 
the substitution $y = {\mathscr B}(z)$ in the inner integral, we obtain
\begin{multline}
  \notag
  \frac{1}{2} \int_{\mathbb R} \d z \int_{\text{SL}(2,\mathbb R)}
  \kappa(\d B)\, 
  \ln \left [ 1 + {\mathscr B}(z)^2 \right ] \, f(z)  \\
=  \frac{1}{2} \int_{\text{SL}(2,\mathbb R)} \kappa (\d B)\,
\int_{\mathbb R}\d z\,  \ln \left [ 1 + {\mathscr B}(z)^2 \right ] f(z) \\
= \frac{1}{2} \int_{\text{SL}(2,\mathbb R)} \kappa (\d B)\, 
\int_{\mathbb R}\d y  \, \ln \left [ 1 + y^2 \right ] 
f \left ( {\mathscr B}^{-1}(y) \right )
\frac{\d {\mathscr B}^{-1}}{\d y} (y) \,.
\end{multline}
Next, we use the letter $z$ instead of $y$, and change the order of integration again: Equation (\ref{lyapunovEquality})
becomes
\begin{multline}
  \notag
  \gamma_\mu = {\mathbb E} \left ( \ln | b_{22} | \right ) \\
  + \frac{1}{2} \int_{\mathbb R} \, \d z  
  \int_{\text{SL}(2,\mathbb R)}\kappa (\d B)\, 
  \left[ f \left( {\mathscr B}^{-1}(z) \right)\, 
    \frac{\d {\mathscr B}^{-1}}{\d z}(z)  -  f(z)
  \right] \,
  \ln (1+z^2)\,. 
\end{multline}
Finally, by making use of Equation
(\ref{frischLloydEquationInDifferentialForm}), and then integrating by
parts, we arrive at the following formula: 
\begin{equation}
  \gamma_\mu = \frac{1}{p} \, \gamma 
  \label{simplifiedLyapunovFormula}
\end{equation}
where
\begin{equation}
  \gamma  := p\,{\mathbb E} \left ( \ln | b_{22} | \right ) 
  + \dashint_{-\infty}^{\infty}\d z\, z f(z)\,, \quad b_{21}=0\,.
\label{reducedLyapunovForUpperTriangularB}
\end{equation}
This formula remains unchanged after restoring $k$ by rescaling.
A similar calculation may be carried out if, instead, $B$ is lower triangular. Equation (\ref{simplifiedLyapunovFormula}) then holds with
\begin{equation}
  \gamma  := p\,{\mathbb E} \left ( \ln | b_{11} | \right ) 
  -  \dashint_{-\infty}^{\infty}\d z\, \frac{E}{z} f(z)\,, \quad b_{12} =0\,.
\label{reducedLyapunovForLowerTriangularB}
\end{equation}
The integrals in these expressions are Cauchy principal value
integrals.  

We shall henceforth refer to $\gamma$ as the {\em reduced Lyapunov exponent}.
Although our derivation of the relation between $\gamma_\mu$ and $\gamma$ assumed
that $E>0$, we conjecture, on the basis of the numerical evidence obtained in all the examples we considered,
that it holds also when $E<0$.

Such simplified formulae for 
the Lyapunov exponent are well-known in the physics
literature \cite{LGP}. 
The reduced
Lyapunov exponent is the rate of growth of the solution of the Schr\"{o}dinger equation:
$$
\gamma = \lim_{x \rightarrow \infty} \frac{1}{x} \ln \sqrt{\psi(x)^2 + \left [ \psi'(x) \right ]^2 }\,.
$$
Alternatively, using the stationarity of the process $\{z(x)\}$,
\begin{equation}
\notag
\gamma 
= \lim_{x \rightarrow \infty} \frac{1}{x} \ln \left | \psi(x)\right | + \lim_{x \rightarrow \infty} \frac{1}{x} \ln \sqrt{1 + z^2(x) }
= \lim_{x \rightarrow \infty} \frac{1}{x} \ln \left | \psi(x)\right |
\end{equation}
and
\begin{equation}
\notag
\gamma 
= \lim_{x \rightarrow \infty} \frac{1}{x} \ln \left | \psi'(x)\right | + \lim_{x \rightarrow \infty} \frac{1}{x} \ln \sqrt{1/z^2(x) + 1 }
= \lim_{x \rightarrow \infty} \frac{1}{x} \ln \left | \psi'(x)\right |\,.
\end{equation}
$\gamma$ also provides a reasonable definition of (the reciprocal of) the {\em localisation length} of the system.
 
The presence of the expectation term on the right-hand side of
Equation \eqref{reducedLyapunovForUpperTriangularB} may, at first sight, surprise
readers familiar with the case of delta scatterers, but its occurence in our more general context is easily explained as follows:
between consecutive scatterers, the wave function is continuous and so, for $x_n < x < x_{n+1}$, we can write
\begin{multline}
\ln |\psi(x) |
=\ln |\psi(x_{n}+) |+\int_{x_{n}}^x \d y \,\frac{\d}{\d y}\ln | \psi(y) | \\
=\ln| \psi(x_{n}+) | + \int_{x_{n}}^x \d y\,z(y)
\,.
\label{logOfPsi}
\end{multline}
Let us denote by $b_{ij}^{(n)}$ the entry of $B_n$ in the $i$th row
and $j$th column. 
If $B$ is upper triangular, we have, at $x_n$,
$$
\psi(x_n+) = b_{22}^{(n)} \,\psi(x_n-)
$$
and so the wavefunction is discontinuous there unless $b_{22}^{(n)} = 1$.
Reporting this in (\ref{logOfPsi}) and iterating, we obtain
$$
\ln|\psi(x)|
= \ln|\psi(0)| + \sum_{j=1}^{n(x)} \ln \left | b_{22}^{(j)} \right |
 + \int_0^x \d y\,z(y)
$$
where $n(x)$, as defined in \S \ref{frischLloydEquationSubsection}, is
the number of point scatterers in 
the interval $[0,x]$. Using
$$
{\mathbb E}(n(x))=px
$$
and the ergodicity of the Riccati 
variable, we recover by this other route 
the equation \eqref{reducedLyapunovForUpperTriangularB} obtained earlier. It is now clear that the expectation term arises
from the possible discontinuities of the wave function at the scatterers.
To give two examples: for the delta scatterer, the wave function is continuous everywhere, $b_{22} = 1$
and
so Equation (\ref{reducedLyapunovForUpperTriangularB}) is just the familiar formula in \cite{LGP}. For the
supersymmetric
scatterer, however, the wavefunction has discontinuites,
$$
b_{22} = \e^{-w}
$$
and so, as noted in \cite{TexHag09}, the formula for the reduced Lyapunov exponent must include the additional term
$$
{\mathbb E}\left (\ln|b_{22}| \right )=-{\mathbb E}(w)\,.
$$

If $B$ is lower triangular instead, it is more natural to work with $\psi'$:
for $x_n < x < x_{n+1}$, we have
\begin{multline}
\notag
\ln |\psi'(x) |
=\ln |\psi'(x_{n}+) |+\int_{x_{n}}^x \d y \,\frac{\d}{\d y}\ln | \psi'(y) | \\
=\ln| \psi'(x_{n}+) | + \int_{x_n}^ x \frac{\psi''(y)}{\psi'(y)}\,\d y = \ln| \psi'(x_{n}+) | -k^2 \int_{x_n}^ x \frac{\psi(y)}{\psi'(y)}\,\d y \\
= \ln| \psi'(x_{n}+) | - k^2 \int_{x_{n}}^x \d y\,\frac{1}{z(y)}
\,.
\end{multline}
Using the lower triangularity of $B_n$, we obtain, at $x_n$,
$$
\psi'(x_n+) = b_{11}^{(n)} \,\psi'(x_n-)\,.
$$
By repeating our earlier argument, we recover Equation (\ref{reducedLyapunovForLowerTriangularB}).

\subsection{Halperin's trick and the energy parameter}
\label{halperinSubsection}
For the particular case of delta scatterers,
Halperin \cite{Ha} devised an ingenious method that, at least in some cases, by-passes the need for quadrature and
yields analytical expressions for the reduced Lyapunov exponent. Let us give a brief outline
of Halperin's trick and discuss some of its consequences.

Halperin works with the Fourier transform of the invariant density:
\begin{equation}
F (x) :=  \int_{\mathbb R} f(z) \,\e^{-\i x z } \,\d z\,.
\label{fourierTransform}
\end{equation}

For the delta scatterer, the Frisch--Lloyd equation (\ref{frischLloydEquation}) in Fourier space is then
\begin{equation}
  F''(x) - E \,F(x) 
  -p  \,\frac{{\mathbb E} \left ( \e^{-\i x u} \right ) -1}{\i x} \,
  F(x) 
  = - 2 \pi \, N \,\delta(x)\,.
  \label{EquationForDeltaScatterers}
\end{equation} 

Let $\varepsilon>0$ and integrate over the interval $(-\varepsilon,\varepsilon)$. Using the fact that
$$
F' (-\varepsilon) =  -\overline{F' (\varepsilon)}
$$
and letting $\varepsilon \rightarrow 0+$, we obtain
$$
  N = -\frac{1}{\pi} \text{Re} \left [ F'(0+) \right ]
  \:.
$$
Furthermore, since in this case $b_{11}=1$ and $b_{21}=0$, Equation
(\ref{reducedLyapunovForUpperTriangularB}) leads to
$$
  \gamma = 
  \dashint_{-\infty}^{\infty} z\, f(z) \,\d z 
  = - \text{Im} \left[ F'(0+) \right]\,.
$$

These two formulae may be combined neatly by introducing the so-called
{\em characteristic function} $\Omega$ associated with the system
\cite{Lu,Nie82}: 
\begin{equation}
  \Omega (E) :=  \gamma(E) - \i \pi N(E) \,.
  \label{characteristicFunction}
\end{equation}
Then Halperin observes that
\begin{equation}
\Omega (E) = \i \frac{F'(0+)}{F(0+)}
\label{halperinFormula}
\end{equation}
where, with a slight abuse of notation, $F$ is now the particular solution of the {\em homogeneous version} of Equation (\ref{EquationForDeltaScatterers})
satisfying the
condition
$$
\lim_{x \rightarrow +\infty} F(x) = 0\,.
$$
Thus the problem of evaluating $\gamma$ and $N$ has been reduced to that of finding the recessive solution of a 
linear homogeneous differential
equation.

Equation (\ref{halperinFormula}) expresses a relationship between the density of states and the Lyapunov exponent--- a relationship
made more explicit in the {\em Herbert--Jones--Thouless
formula} \cite{HJ,Th} well-known in the theory of quantum disordered systems. 
A further consequence of the same equation is that, if
the recessive 
solution $F$ depends analytically on the energy parameter, so does the
characteristic function. $\Omega$ should thus have an analytic continuation everywhere in the
complex plane, save on the cut where the essential
spectrum of the Schr\"{o}dinger Hamiltonian lies. 

More generally, for an arbitrary scatterer, the Fourier transform $F$
of the invariant density satisfies the equation
\begin{multline}
F''(x) - E \,F(x) 
- p \int_{\text{SL} (2,\mathbb R)} \kappa (\d B) \int_{\mathbb R} \d z\,\frac{\e^{-\i x {\mathscr B}(z)}-\e^{-\i x z}}{\i x} \, f(z) 
= 0\,.
\label{frischLloydEquationInFourierSpace}
\end{multline} 
We shall not make explicit use of this equation in what follows: instead,
we shall obtain closed formulae for the characteristic function by making use of analytic continuation.
This is one important benefit of having retained the energy parameter.

\subsection{Random continued fractions}
\label{continuedFractionSubsection}
There is a close correspondence between products of $2 \times 2$
matrices and continued fractions. 
Let $z_0$ be an arbitrary starting value, recall the definition (\ref{action}) and set
\begin{equation}
z_{n} := {\mathscr A}_{n-1} \circ \cdots \circ {\mathscr A}_0 (z_0)\,.
\label{forwardIteration}
\end{equation}
Then the sequence $\{ z_n \}_{n \in {\mathbb N}}$ is a Markov chain on the projective line, and $\nu$ is $\mu$-invariant if and only if it is
a stationary distribution of this Markov chain. Now,
reverse the order of the matrices in the product, set $\zeta_0 = z_0$ and
\begin{equation}
\zeta_{n}  := {\mathscr A}_0 \circ \cdots \circ {\mathscr A}_{n-1} (\zeta_0)\,.
\label{backwardIteration}
\end{equation}
Although, for every $n$, $z_n$ and $\zeta_n$ have the same distribution, the large-$n$ behaviour of a typical realisation of the sequence
$\{z_n\}_{n \in {\mathbb N}}$ differs greatly from that of a typical realisation of the sequence $\{\zeta_n\}_{n \in {\mathbb N}}$ \cite{Fu}.

$\{\zeta_n\}_{n \in {\mathbb N}}$ converges to a (random) limit, say
$\zeta$. Write 
$$
A_n := \begin{pmatrix}
a_n & b_n \\
c_n & d_n
\end{pmatrix}\,.
$$
Then
\begin{equation}
{\mathscr A}_n (\zeta) = a_n/c_n - \frac{1/c_n^2}{d_n/c_n + \zeta}
\label{continuedFractionForm}
\end{equation}
and so
\begin{equation}
\zeta := a_0/c_0 - \cfrac{1/c_0^2}{d_0/c_0 + a_1/c_1 - \cfrac{1/c_1^2}{d_1/c_1 + a_2/c_2 - \cfrac{1/c_2^2}{d_2/c_2 + \cdots}}}\,.
\label{continuedFraction}
\end{equation}
It is immediately clear that, if $A$ is $\mu$-distributed, then
$$ 
{\mathscr A} (\zeta) = a/c - \frac{1/c^2}{d/c + \zeta}
$$
has the same distribution as $\zeta$. Hence the distribution of $\zeta$ is $\mu$-invariant. Furthermore, if $\zeta$ 
is independent of $z_0$, then there can be only one $\mu$-invariant measure. So $f$ is also the density of the
infinite random continued fraction $\zeta$.

By contrast,
$\{z_n\}_{n \in {\mathbb N}}$ behaves ergodically. Therefore the density $f$ of the invariant measure $\nu$ should
be well approximated by a histogram of the $z_n$. We have used this to verify
the correctness of our results. 

\section{Some explicit invariant measures}
\label{invariantSection}

\subsection{Delta scatterers}
\label{deltaSubsection}
In this section, we obtain invariant measures for products where the
matrices are of the form 
\begin{equation}
A = 
\begin{pmatrix}
\sqrt{k} & 0 \\
0 & \frac{1}{\sqrt{k}}
\end{pmatrix}
\begin{pmatrix}
\cos (k\theta)  & -\sin (k\theta)  \\
\sin (k\theta) & \cos (k\theta )
\end{pmatrix} 
\begin{pmatrix}
\frac{1}{\sqrt{k}} & 0 \\
0 & \sqrt{k}
\end{pmatrix}
\begin{pmatrix}
1 & u \\
0 & 1
\end{pmatrix}
\label{deltaInteractionForPositiveEnergy}
\end{equation}
or
\begin{equation}
A = 
\begin{pmatrix}
\sqrt{k} & 0 \\
0 & \frac{1}{\sqrt{k}}
\end{pmatrix}
\begin{pmatrix}
\cosh  (k \theta)  & \sinh  (k\theta) \\
\sinh (k \theta)  & \cosh (k\theta) 
\end{pmatrix} 
\begin{pmatrix}
\frac{1}{\sqrt{k}} & 0 \\
0 & \sqrt{k}
\end{pmatrix}
\begin{pmatrix}
1 & u \\
0 & 1
\end{pmatrix}
\label{deltaInteractionForNegativeEnergy}
\end{equation}
where $\theta \sim \text{Exp}(p)$ and $u$ is a random variable,
independent of $\theta$, whose density we denote by 
$\varrho : {\mathbb R} \rightarrow {\mathbb R}_+$. These products are
associated with the generalised Kronig--Penney model 
for $E = k^2 >0$ and $E = -k^2$ respectively, in the case where (see
Example \ref{deltaExample}) 
$$
B = \begin{pmatrix}
1 & u \\
0 & 1
\end{pmatrix}\,.
$$

The corresponding Frisch--Lloyd equation (\ref{frischLloydEquation}) is
$$
(z^2 + E)\,f(z) + p \int_{\mathbb R} \d u \int_{z}^{z-u}\d t  \,  
f(t)\,\varrho ( u)  =N\,.
$$
We change the order of integration; the equation becomes
\begin{equation}
  N = (z^2 + E)\,f(z) + p \int_{\mathbb R} \d t\,  K (z-t)\,f(t) 
  \label{frischLloydForDelta}
\end{equation}
where
\begin{equation}
K(x) = \begin{cases}
- \int_x^\infty \varrho(u) \,\d u & \text{if $x > 0$} \\
\int_{-\infty}^x \varrho(u)\,\d u & \text{if $ x<0$}
\end{cases}\,.
\label{deltaKernel}
\end{equation}

Suppose that
\begin{equation}
\pm u \sim \text{\tt Exp}(q)\,.
\label{exponentialDistributionForTheDeltaInteraction}
\end{equation}
We shall show that, in this case, the Frisch--Lloyd equation reduces to a first-order differential equation. For
the sake of clarity, consider first the case $u \sim \text{Exp} (q)$.
For this choice of distribution,
$$
K(x) = \begin{cases}
- \e^{-q x} & \text{if $x>0$} \\
0 & \text{if $x<0$} 
\end{cases}\,.
$$
So we have
$$
K'(x) = - q K(x)\,, \;\; K(0+) = -1\,,
$$
and equation (\ref{frischLloydForDelta}) is
$$
N = (z^2+E)\, f(z) + p \int_{-\infty}^z\d t \, K(z-t)\,f(t)\,.
$$
Differentiate this equation with respect to $z$:
\begin{multline}
\notag
0 = \frac{\d}{\d z} \left [ (z^2+E)\,f(z) \right ] + p K(0+) f(z) + p
\int_{-\infty}^z\d t \, K'(z-t)\,f(t) \\
=  \frac{\d}{\d z} \left [ (z^2+E)\,f(z) \right ] - p f(z) - q p
\int_{-\infty}^z\d t \, K(z-t)\,f(t) \\
= \frac{\d}{\d z} \left [ (z^2+E)\,f(z) \right ] - p f(z) - q \left [ N - (z^2+E)\,f(z) \right ]\,.
\end{multline}
This is the required differential equation. The case $-u \sim \text{Exp}(q)$ is analogous, 
and so we find, for the general case
(\ref{exponentialDistributionForTheDeltaInteraction}), 
\begin{equation}
  \frac{\d}{\d z} \left [ (z^2 + E) \,f(z) \right ] 
  - p \,f(z) \pm q \left [ (z^2 + E) \,f(z) \right ] = \pm q N\,.
  \label{deltaDifferentialEquationForExponential}
\end{equation}

We seek the particular solution that satisfies
the normalisation condition
\begin{equation}
  \int_{\mathbb R} f(z) \,\d z = 1\,.
  \label{normalisationCondition}
\end{equation}
This condition fixes the constant of integration $N$,
and hence provides an expression for the integrated density of states
for the Schr\"odinger Hamiltonian.

\subsubsection{Product of the form (\ref{deltaInteractionForPositiveEnergy})}
For $E = k^2 > 0$, this leads to
\begin{multline}
  f(z) =  
  \frac{\pm q N}{z^2+k^2}
  \exp \left[ \mp q z + \frac{p}{k} \arctan \frac{z}{k} \right]  \\
  \times \int_{\mp \infty}^z 
  \exp \left[ \pm q t - \frac{p}{k} \arctan \frac{t}{k} \right]\,\d t\,.
\label{nuForDeltaInteractionWithExponentialStrengthandPositiveE}
\end{multline}

\begin{figure}[htbp]
  \centering
  \includegraphics[scale=0.725]{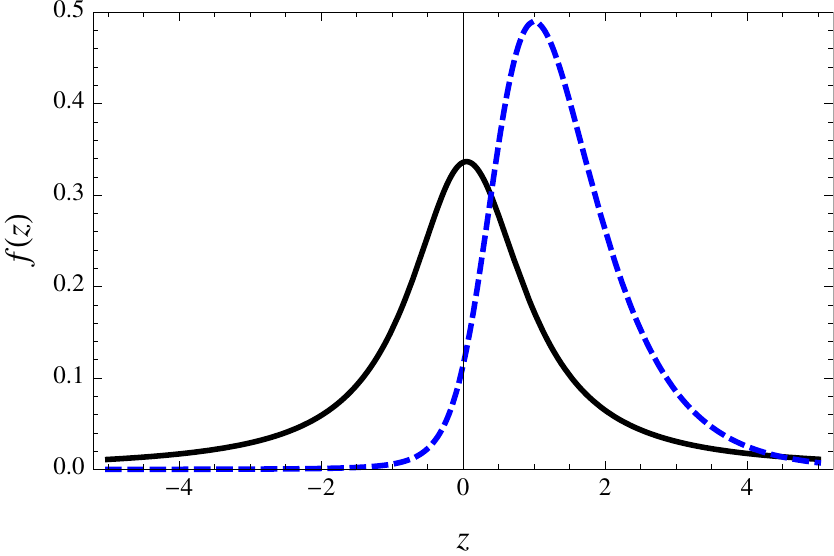}
  \hspace{0.1cm}
  \includegraphics[scale=0.725]{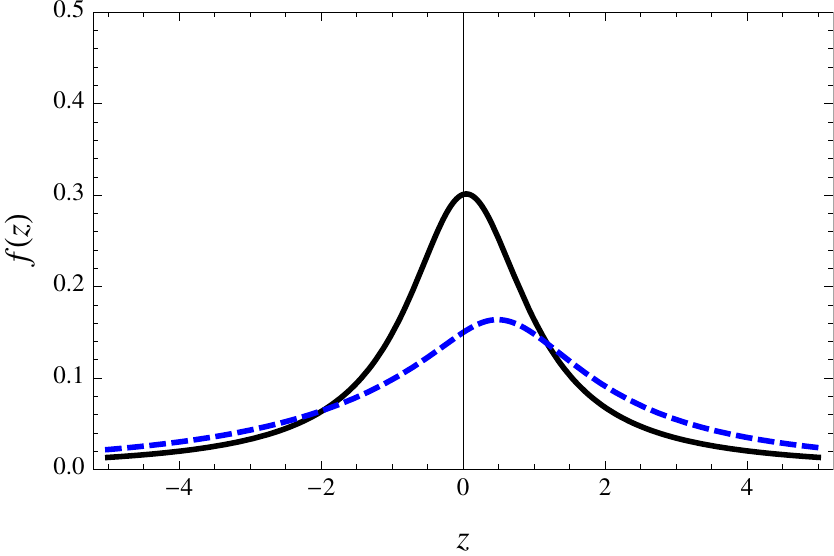}
  \vspace{0.1cm}
  \caption{Plots of the invariant density $f$ for delta scatterers and
    positive energy $E=k^2=+1$.
    Black continuous lines correspond to a low density of scatterers, i.e.
    $\theta\sim{\tt Exp}(p)$ with $p=1/4$, and blue dashed lines to a
    high density, i.e. $p=4$.
    (a) $u\sim\text{\tt Exp}(1)$;
    (b) $-u\sim\text{\tt Exp}(1)$.}
    \begin{picture}(0,0) 
         \put(-85,80){(a)}
         \put(95,80){(b)}
    \end{picture}
  \label{fig:fdelta_pe}
\end{figure}
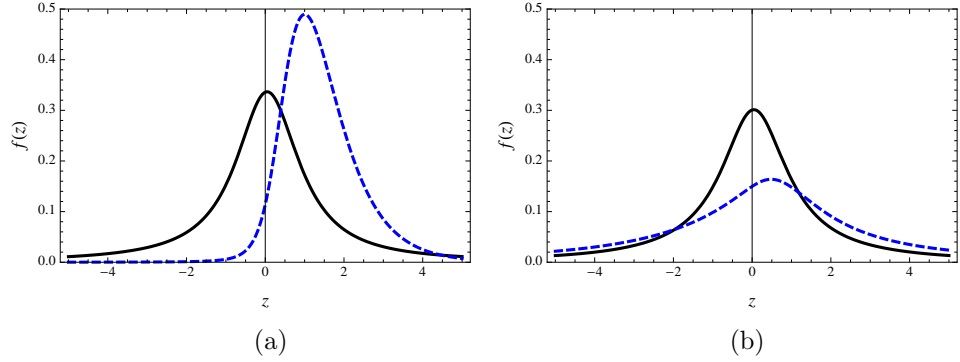

The density of the Riccati variable is plotted
  in Figure \ref{fig:fdelta_pe} for (a) positive $u_n$ and
  (b) negative $u_n$. 
  The continuous black curves correspond to a low density of
  scatterers (small $p$, compared to $k$ and $1/q$) 
  and are reminiscent of the Cauchy law
  obtained in the absence of scatterers. Recall that the effect of the $n$th scatterer on the
  Riccati variable is described by the equation
  $$
  z(x_n+)=z(x_n-)+u_n\,.
  $$
  When the $u_n$ are positive, any increase in  
  the concentration of the scatterers produces a
  decrease in the current and so 
  the distribution is pushed to the right; see the blue dashed curve in Figure \ref{fig:fdelta_pe} (a).
  On the other hand, when the $u_n$ are negative, any increase in the 
  concentration of the scatterers 
  leads to an increase in the current of the Riccati variable and so spreads the
  distribution; see the blue dashed curve in Figure \ref{fig:fdelta_pe} (b).

\subsubsection{Product of the form (\ref{deltaInteractionForNegativeEnergy})}
For $E = -k^2 < 0$ and $u \sim \text{Exp} (q)$, one must take
$N=0$ to obtain a normalisable solution--- a reflection of the fact that the
essential spectrum of the Schr\"odinger  Hamiltonian is 
${\mathbb  R}_+$. 
Then 
\begin{equation}
  f(z) = C^{-1}
  \frac{\e^{-q z}}{z^2-k^2} 
  \left ( \frac{z-k}{z+k} \right )^{\frac{p}{2k}} {\mathbf 1}_{(k,\infty)}(z)
  \label{nuForFLandNegativeEplus}
\end{equation}
where $C$ is the normalisation constant.

By contrast, in the case $-u \sim \text{\tt Exp}(q)$, one finds
\begin{equation}
f(z) = \frac{q N}{z^2-k^2} \e^{q z}
\left | \frac{z-k}{z+k} \right |^{\frac{p}{2k}} \int_z^{c(z)} \e^{-q t} \left | \frac{t+k}{t-k} \right |^{\frac{p}{2k}}\,\d t
\label{nuForFLandNegativeEminus}
\end{equation}
where
$$
  c(z) = \begin{cases}
    \infty & \text{if $z > k$} \\
    -k & \text{if $z < k$}
  \end{cases}
  \:.
$$

\begin{figure}[htbp]
  \centering
  \includegraphics[scale=0.725]{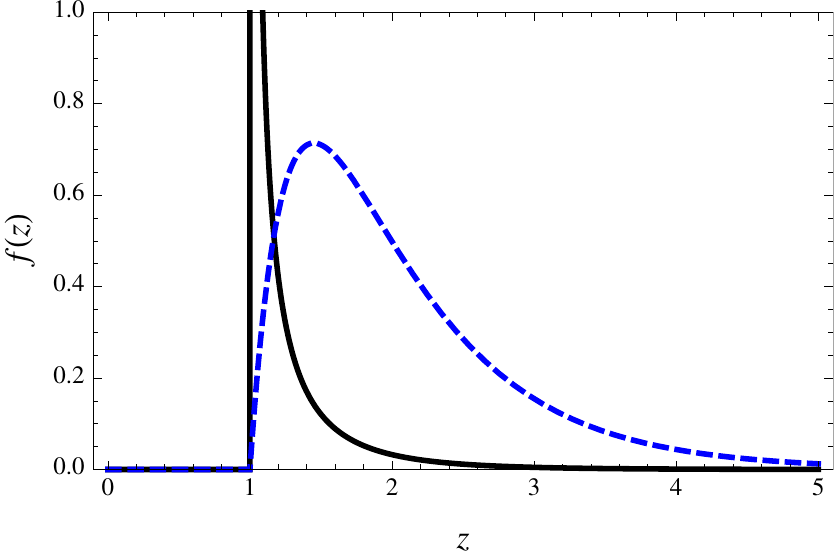}
  \hspace{0.1cm}
  \includegraphics[scale=0.725]{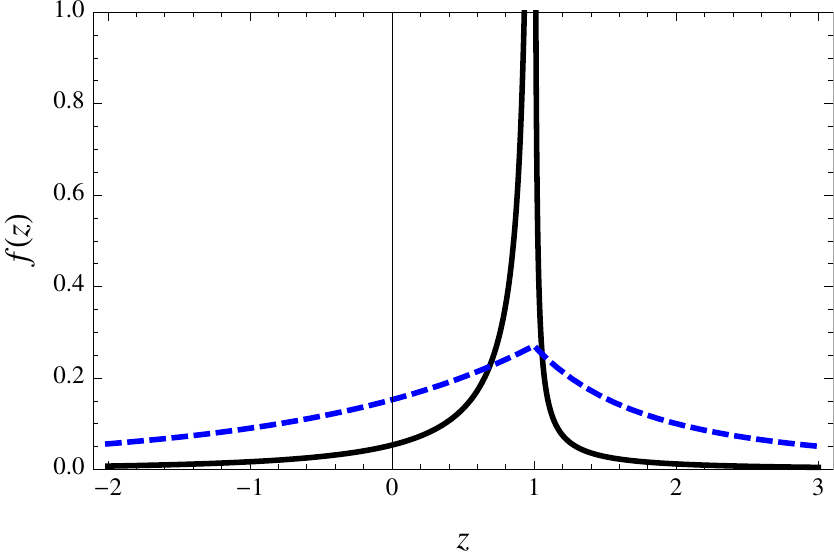}
  \vspace{0.1cm}
  \caption{Plots of the invariant density $f$  for delta scatterers and
    negative energy $E=-k^2=-1$.
    Black continuous lines correspond to a ``low'' density of scatterers, i.e.
    $\theta\sim{\tt Exp}(p)$ with $p=1/4$, and blue dashed lines to a
    ``high'' density $p=4$.
    (a) $u\sim\text{\tt Exp}(1)$;
    (b) $-u\sim\text{\tt Exp}(1)$.}
    \begin{picture}(0,0) 
         \put(-85,80){(a)}
         \put(95,80){(b)}
    \end{picture}
  \label{fig:fdelta_ne}
\end{figure}
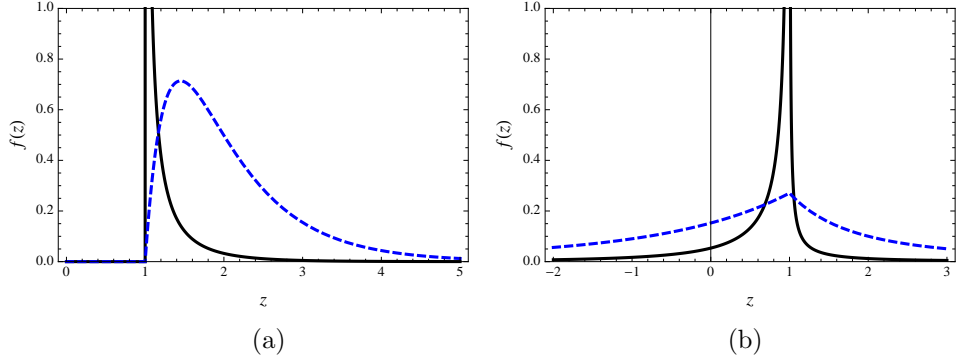

The invariant density $f$ is plotted
  in Figure \ref{fig:fdelta_ne} for the cases (a) $u \sim {\tt Exp}(q)$
  and (b) $-u \sim {\tt Exp}(q)$ respectively.
  The shape of the distribution can again be explained by using
  the qualitative picture of \S \ref{qualitativeSubsection}.
  For positive $u_n$, the sharp peak obtained for 
  a small concentration $p$ of scatterers (black continuous line)
  reflects the trapping of 
  the process $\{ z(x)\}$ by the potential well at $z=k$; recall Figure~\ref{fig:pot_U}. 
  When the concentration of scatterers is increased, the Riccati
  variable experiences positive jumps more frequently
  and so the distribution spreads to the right
  (blue dashed curve).  
For negative $u_n$, the jumps
can take arbitrary negative values. This enables the ``particle'' to
overcome the barrier at $z=-k$, and so we have a non-zero current $N$
(i.e. a non-zero density of states).
This effect is enhanced as the density of the scatterers is
increased (blue dashed curve).

\subsubsection{Calculation of the characteristic function} Using the invariant measure, it is trivial to express the integrated density of states 
and the Lyapunov exponent in integral form. Such integral expressions are particularly useful when studying the asymptotics of $N$ and $\gamma$
in various limits. Even so, it is worth seeking analytical expressions (in terms of special functions) for these quantities, as
they sometimes reveal unexpected connections to other problems. 

Recalling the discussion in \S \ref{halperinSubsection}, we begin with
a straightforward application of Halperin's trick. For $\pm u \sim {\tt Exp}(q)$, we have
$$
{\mathbb E} \left ( \e^{-\i u x } \right ) = \frac{1}{1 \pm \i x/q}
$$
and so the homogeneous version of Equation (\ref{EquationForDeltaScatterers}) is
\begin{equation}
\notag
F''(x) + \left [ -E  + \frac{p}{\pm q + \i x} \right ] \,F(x) = 0\,.
\end{equation} 
The recessive solution is
$$
F(x) = \frac{W_{\frac{-p}{2 \sqrt{-E}},\frac{1}{2}}\left ( 2 \sqrt{-E} \left [ \i x \pm q \right ] \right )}{W_{\frac{-p}{2 \sqrt{-E}},\frac{1}{2}}\left ( \pm 2 \sqrt{-E}  q \right )}
$$
where $W_{a,b}$ is the Whittaker function~\cite{AS,GR}.
We deduce that, for $E$ outside the essential spectrum of the
Schr\"{o}dinger Hamiltonian, 
\begin{equation}
  \Omega (E) := \gamma (E) -\i \pi N (E) 
  = - 2 \sqrt{-E} \, 
  \frac{W_{\frac{-p}{2 \sqrt{-E}},\frac{1}{2}}'\left(\pm2 \sqrt{-E} q \right)}
       {W_{\frac{-p}{2 \sqrt{-E}},\frac{1}{2}}\left(\pm2 \sqrt{-E} q \right)}\,.
  \label{characteristicFunctionForDeltaScatterer}
\end{equation}
This formula for the characteristic function was discovered by
Nieuwenhuizen \cite{Ni}. In particular, for $k$ real,
$$
\gamma (k^2) - \i \pi N (k^2) = \Omega (k^2+\i 0+)
= 2 \i k \frac{W_{\frac{-\i p}{2 k},\frac{1}{2}} ' \left (  \mp 2 \i k q \right )}{W_{\frac{-\i p}{2 k},\frac{1}{2}}\left ( \mp 2 \i k  q \right )}
$$
and
$$
\gamma (-k^2) - \i \pi N (-k^2) = \Omega (-k^2+\i 0+)
= - 2 k \frac{W_{\frac{-p}{2 k},\frac{1}{2}} ' \left (  \pm 2 k q \right )}{W_{\frac{- p}{2 k},\frac{1}{2}}\left ( \pm 2 k  q \right )}\,.
$$

In the case $u \sim {\tt Exp}(q)$ there is an alternative derivation of this formula which does not require the solution of a differential equation: start with the
explicit form  of the invariant density $f$ for $E=-k^2<0$, given by Equation (\ref{nuForFLandNegativeEplus}). By using Formula 3 in \cite{GR}, \S 3.384, we obtain the following expression for the normalisation constant:
$$
C := \int_k^{\infty} \frac{\e^{-q z}}{z^2-k^2} 
  \left ( \frac{z-k}{z+k} \right )^{\frac{p}{2k}} \,\d z = \frac{1}{2 k} \Gamma \left ( \frac{p}{2 k} \right )\,
  W_{\frac{-p}{2 k},\frac{1}{2}} \left ( 2 k q \right )\,.
$$
The reduced Lyapunov exponent $\gamma$ may then be obtained easily by noticing that 
differentiation with respect to the parameter $q$ yields an additional factor of
$z$ in the integrand. Hence, for $k$ real,
we find
\begin{equation*}
   \gamma (-k^2) =  \int_{k}^{\infty}\d z\, z \, f(z)
  = -\frac{\partial}{\partial q}\ln C
  = -2  k\, \frac{W'_{\frac{-p}{2 k},\frac{1}{2}} \left ( 2 k q \right )}
                {W_{\frac{-p}{2 k},\frac{1}{2}} \left ( 2 k q \right )}
  \,.
\end{equation*}
Since $N=0$ for $E<0$, this yields
\begin{equation}
\Omega(-k^2) = -2  k\, \frac{W'_{\frac{-p}{2 k},\frac{1}{2}} \left ( 2 k q \right )}
                {W_{\frac{-p}{2 k},\frac{1}{2}} \left ( 2 k q \right )}\,.
  \label{eq:LyapunovFrischLloydNegativeEnergy}
\end{equation}
Now,
the half-line $E<0$ lies outside the essential spectrum of the Schr\"{o}dinger Hamiltonian because $N=0$ along it. Hence $\Omega$
is analytic along this half-line, and we see that the ``$+$ case'' of our earlier Equation
(\ref{characteristicFunctionForDeltaScatterer}) is simply the analytic continuation of Equation (\ref{eq:LyapunovFrischLloydNegativeEnergy}).
In particular, the formula in the case $E = k^2>0$ may be deduced from the formula
in the case $E = -k^2 < 0$ by applying the simple substitution
$$
k \mapsto - \i k\,.
$$

\subsection{Delta--prime scatterers}
\label{deltaPrimeSubsection}
Products of matrices of the form
\begin{equation}
  A = 
  \begin{pmatrix}
    \sqrt{k}  & 0 \\
    0 &  \frac{1}{\sqrt{k}}
  \end{pmatrix}
  \begin{pmatrix}
    \cos (k\theta)  & -\sin (k\theta)  \\
    \sin (k\theta)  & \cos  (k \theta) 
  \end{pmatrix} 
  \begin{pmatrix}
     \frac{1}{\sqrt{k}} & 0 \\
     0 & \sqrt{k}
  \end{pmatrix}
  \begin{pmatrix}
     1 & 0 \\
     v & 1
  \end{pmatrix}
  \label{deltaPrimeInteractionForPositiveEnergy}
\end{equation}
or
\begin{equation}
  A = 
  \begin{pmatrix}
     \sqrt{k}  & 0 \\
     0 &  \frac{1}{\sqrt{k}}
  \end{pmatrix}
  \begin{pmatrix}
     \cosh (k\theta)  & \sinh  (k\theta)  \\
     \sinh (k \theta)  & \cosh (k\theta) 
  \end{pmatrix} 
  \begin{pmatrix}
     \frac{1}{\sqrt{k}} & 0 \\
     0 & \sqrt{k}
  \end{pmatrix}
  \begin{pmatrix}
     1 & 0 \\
     v & 1
  \end{pmatrix}
  \label{deltaPrimeInteractionForNegativeEnergy}
\end{equation}
where $\theta \sim \text{Exp}(p)$ and $v$ is a random variable independent of $\theta$, are associated with
the delta-prime scatterer (see Example \ref{deltaPrimeExample})
$$
B = \begin{pmatrix}
1 & 0 \\
v & 1
\end{pmatrix}\,.
$$

The Frisch--Lloyd equation (\ref{frischLloydEquation}) for this scatterer is
\begin{equation}
  (z^2 + E)\,f(z) 
  + p \int_{\mathbb R} \d v \int_{z}^{\frac{z}{1-vz}}\d t\,  f(t)\,\varrho(v) =N
  \label{frischLloydForDeltaPrime}
\end{equation}
where $\varrho$ is the density of $v$.
The calculation of the invariant measure in this case can be reduced to the calculation of the invariant measure for some 
Kronig--Penney model with delta scatterers. For instance, in the negative energy case (\ref{deltaPrimeInteractionForNegativeEnergy}) with $k=1$,
we have
$$
\begin{pmatrix}
  0 & 1 \\ 1 & 0
\end{pmatrix}
A
\begin{pmatrix}
  0 & 1 \\ 1 & 0
\end{pmatrix}
=
  \begin{pmatrix}
    \cosh \theta  &  \sinh \theta  \\
    \sinh \theta  &  \cosh \theta 
  \end{pmatrix} 
  \begin{pmatrix}
     1 & v \\
     0 & 1
  \end{pmatrix}
\,.
$$
The similarity transformation of the matrix $A$ on the left corresponds to the transformation
$z \mapsto 1/z$ of the Riccati variable. So the invariant densities for the delta and the delta-prime cases are in
a reciprocal relationship.    
Accordingly, replace $z$ by $1/z$ in Equation (\ref{frischLloydForDeltaPrime}) and set
$$
g(z) = \frac{1}{z^2} f(1/z)\,.
$$
Then
\begin{multline}
  N = \left ( 1+Ez^2 \right ) g(z) 
  - p \int_{\mathbb R} \d v \int_{z}^{z-v}\d t \, g(t) \,\varrho(v) \\
= \left ( 1+Ez^2 \right ) g(z) - p \int_{\mathbb R}\d t \,  K(z-t) \,g(t)
\label{frischLloydForDeltaPrime2}
\end{multline}
where $K$ is the kernel defined by Equation (\ref{deltaKernel}). This equation for $g$ is essentially the same
as Equation (\ref{frischLloydForDelta}) save for the sign of $p$ and the dependence on the energy. For the case
$$
\pm v \sim {\tt Exp} (q)
$$
this equation can, by using the same tricks as before, be converted
into a differential equation which is easy to solve.

\subsubsection{Product of the form (\ref{deltaPrimeInteractionForPositiveEnergy})}
For $E = k^2 > 0$, the upshot is
\begin{multline}
  f(z) =  
  \frac{\pm q N}{z^2+k^2}
  \exp \left[ \mp \frac{q}{z} - \frac{p}{k} \arctan \frac{k}{z} \right]  \\
  \times \int_{\mp \infty}^{1/z} 
  \exp \left[ \pm q t + \frac{p}{k} \arctan (k t) \right]\,\d t\,.
\label{nuForDeltaPrimeInteractionWithExponentialStrengthandPositiveE}
\end{multline}

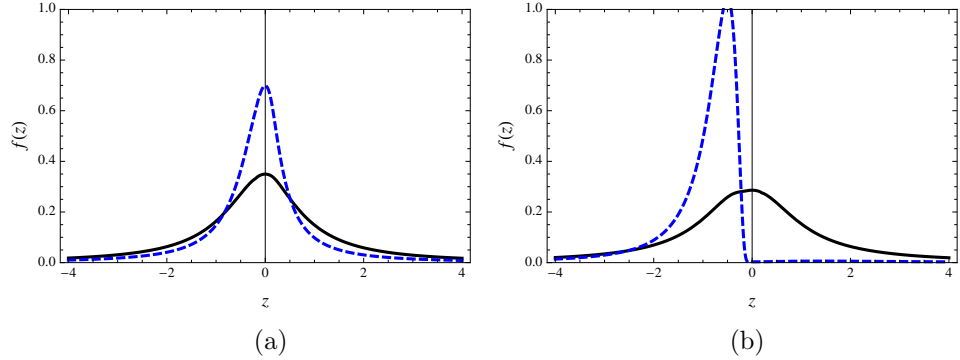
\begin{figure}[htbp]
  \centering
  \includegraphics[scale=0.725]{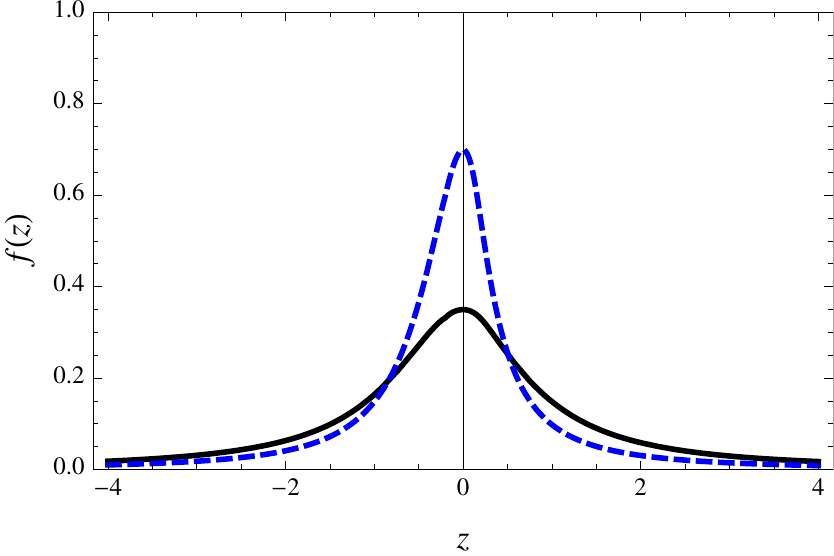}
  \hspace{0.1cm}
  \includegraphics[scale=0.725]{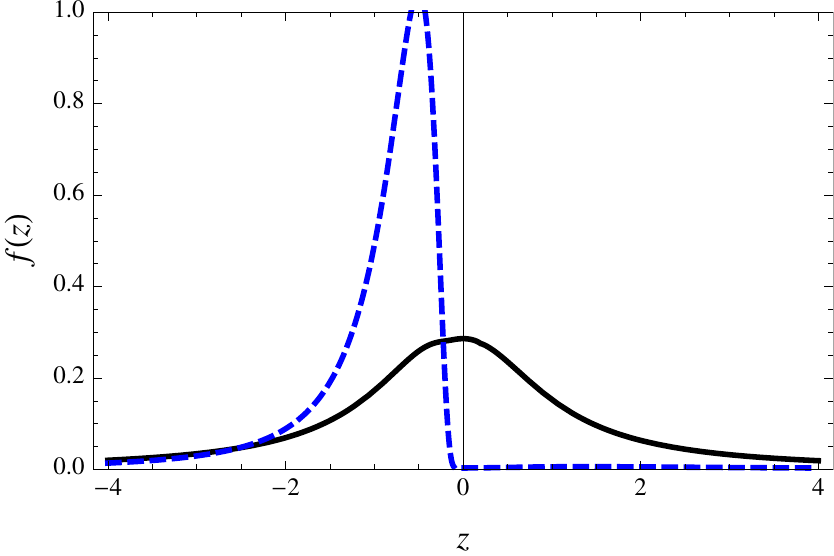}
  \vspace{0.1cm}
  \caption{Plots of the invariant density $f$ for delta-prime scatterers
    and positive energy $E=k^2=+1$.
    Black continuous lines correspond to a ``low'' density of scatterers, i.e.
    $\theta\sim{\tt Exp}(p)$ with $p=1/4$, and blue dashed lines to a
    ``high'' density $p=4$.
    (a) $v\sim\text{\tt Exp}(1)$. 
    (b) $-v\sim\text{\tt Exp}(1)$.}
  \begin{picture}(0,0) 
         \put(-85,80){(a)}
         \put(95,80){(b)}
  \end{picture}
  \label{fig:fdeltaprime_pe}
\end{figure}

Plots of the distribution are shown in Figure \ref{fig:fdeltaprime_pe}
for (a) positive $v_n$ and (b) negative $v_n$. These plots differ
somewhat from those obtained in the case of delta scatterers, and
we can use the particle analogy of \S \ref{qualitativeSubsection} to explain
the differences. The jump of the particle associated with the
$n$th delta-prime scatterer is given implicitly by
\begin{equation}
\frac{1}{z(x_n+)} = \frac{1}{z(x_n-)} + v_n\,.
\label{ricattiJumpForDeltaPrime}
\end{equation}
The strongly asymmetric distribution obtained for negative
$v_n$ (Part (b) of Figure \ref{fig:fdeltaprime_pe}) can be explained as
follows: starting from $+\infty$, the particle experiences its first jump at ``time'' $x_1$, and its
value after the jump is approximately
$1/v_1 < 0$. In fact, for $z$ negative and small in modulus, the invariant density resembles very closely
that of $1/v_1$, i.e.
$$
f(z) \sim \frac{c}{z^2}\e^{q/z} \;\; \text{as $z \rightarrow 0-$}\,.
$$ 
Thereafter, the particle proceeds towards $-\infty$. In particular, if $p$ is large, then the expected value of $x_1$ is small, and the particle
spends hardly any time on the positive semi-axis.

{\em Product of the form (\ref{deltaPrimeInteractionForNegativeEnergy}):}
for $E = -k^2 < 0$ and $v \sim \text{\tt Exp}(q)$, we find
\begin{equation}
  f(z) = C^{-1} \, \frac{\e^{-q/z}}{k^2-z^2} 
  \left( \frac{k-z}{k+z} \right)^{\frac{p}{2 k}} \,{\mathbf 1}_{(0,k)}(z)
\label{densityForDeltaPrimeEnegativeQpositive}
\end{equation}
where
\begin{equation}
  C = \frac{1}{2 k}
           \Gamma\left( \frac{p}{2 k} \right) 
             W_{-\frac{p}{2 k},\frac{1}{2}} \left( 2q/k \right)
            \,.
\label{deltaPrimeNormalisationConstant}            
\end{equation}

When $E=-k^2<0$ and $-v \sim \text{\tt Exp}(q)$, we obtain
\begin{equation}
  f(z) = \frac{q N}{z^2-k^2} \e^{q / z}
  \left| \frac{z-k}{z+k} \right|^{\frac{p}{2k}} 
  \int_{c(z)}^{1/z} \e^{-q t} \left| \frac{1+k t}{1-k t}\right|^{\frac{p}{2k}}
  \,\d t
\label{densityForDeltaPrimeEnegativeQnegative}
\end{equation}
where
$$
c(z) = \begin{cases}
\infty & \text{if $0 < z < 1/k$} \\
-k & \text{otherwise}
\end{cases}
\,.
$$
 
\begin{figure}[htbp]
  \centering
  \includegraphics[scale=0.725]{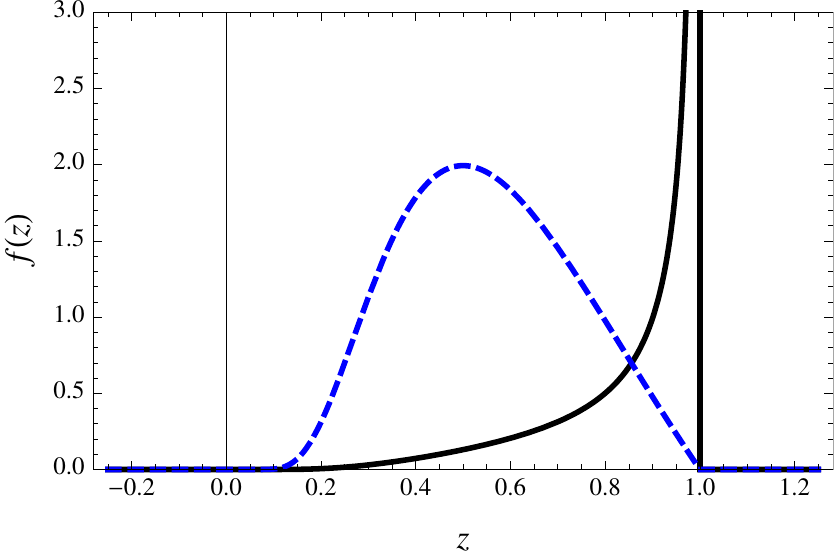}
  \hspace{0.1cm}
  \includegraphics[scale=0.725]{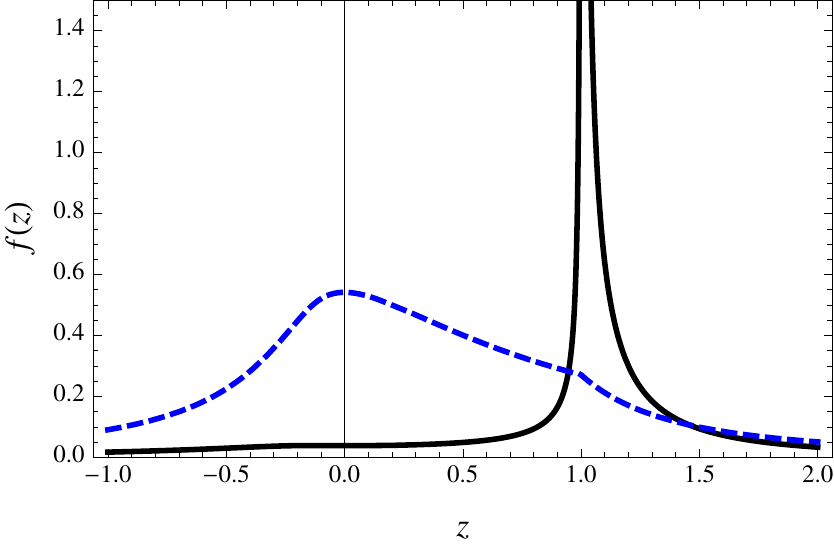}
  \vspace{0.1cm}
  \caption{Plots of the invariant density $f$ for delta-prime scatterers
    and negative energy $E=-k^2=-1$.
    Black continuous lines correspond to a ``low'' density of scatterers, i.e.
    $\theta\sim{\tt Exp}(p)$ with $p=1/4$, and blue dashed lines to a
    ``high'' density $p=4$.
    (a) $v\sim\text{\tt Exp}(1)$. 
    (b) $-v\sim\text{\tt Exp}(1)$.}
  \begin{picture}(0,0) 
         \put(-85,80){(a)}
         \put(95,80){(b)}
  \end{picture}
  \label{fig:fdeltaprime_ne}
\end{figure}
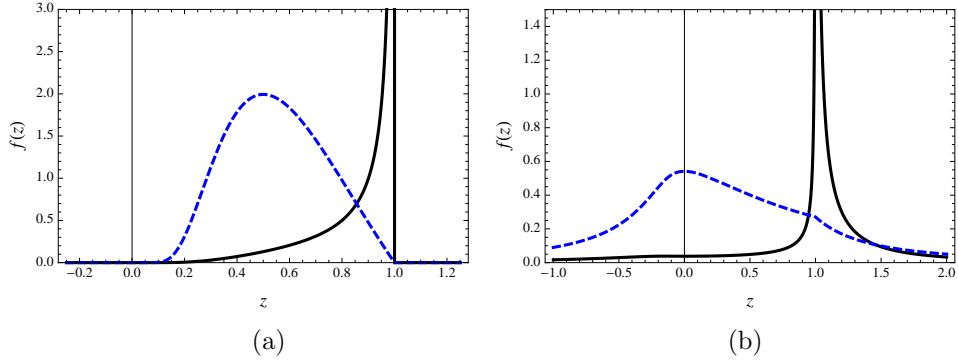

Again, we can try to understand the qualitative features of the density
function $f$ for $E<0$ by invoking the particle analogy of \S
\ref{qualitativeSubsection}. In view of Equation (\ref{ricattiJumpForDeltaPrime}),
when $v_n \sim {\tt Exp}(q)$ and $z(x_n-)>0$, the value of the Riccati
variable decreases but can never become negative. So the particle,
once it passes to the left 
of the equilibrium point at $z=k$, must remain trapped there. This
explains why the density is supported on $(0,k)$; see
Figure \ref{fig:fdeltaprime_ne} (a).
By contrast, when $-v_n \sim {\tt Exp}(q)$, the jumps are unrestricted;
the ``particle'' can escape over the potential barrier at $z=-k$
infinitely often, leading to a non-zero current and a
density $f$ spread over ${\mathbb R}$. This is shown in Figure \ref{fig:fdeltaprime_ne} (b).

\subsubsection{Calculation of the characteristic function}
\label{deltaPrimeCharacteristicSubsection}

We begin with the case $v \sim {\tt Exp}(q)$ and $E= -k^2 < 0$. 

The invariant density is then given
by Equation (\ref{densityForDeltaPrimeEnegativeQpositive}). Using Formula (\ref{reducedLyapunovForLowerTriangularB}) for the
reduced Lyapunov exponent and the expression (\ref{deltaPrimeNormalisationConstant}) for the normalisation constant $C$, we find
$$
\gamma(-k^2) = -k^2 \frac{\partial}{\partial q} \ln C = - 2 k \frac{W_{\frac{-p}{2 k},\frac{1}{2}}'\left ( 2 q/k \right )}{W_{\frac{-p}{2 k},\frac{1}{2}} \left ( 2 q/k \right )}\,.
$$
Since $N(-k^2)=0$ in this case, analytic continuation yields
\begin{equation}
\Omega (E) =  - 2 \sqrt{-E} \,\frac{W_{\frac{-p}{2 \sqrt{-E}},\frac{1}{2}}'\left ( 2 q/\sqrt{-E} \right )}{W_{\frac{-p}{2 \sqrt{-E}},\frac{1}{2}} \left ( 2 q/\sqrt{-E} \right )}\,.
\label{characteristicFunctionForDeltaPrime}
\end{equation}

The characteristic function in the case $-v \sim {\tt Exp}(q)$ is the same, except that $q$ becomes $-q$.
In particular, for $E = k^2 >0$ and $\pm v \sim {\tt Exp}(q)$, we obtain
$$
\gamma (k^2) - \i \pi N (k^2) = \Omega (k^2+\i0+) = 2 \i k \frac{W_{\frac{-\i p}{2 k},\frac{1}{2}}'\left ( \pm 2 \i q/k \right )}{W_{\frac{-\i p}{2 k},\frac{1}{2}} \left ( \pm 2 \i q/k \right )}\,.
$$

An alternative derivation of these results could use the correspondence between the delta and delta-prime cases alluded to earlier.

\subsection{Supersymmetric scatterers}
\label{susySubsection}
We now consider products where the matrices are of the form
\begin{equation}
A = 
\begin{pmatrix}
\sqrt{k}  & 0 \\
0 &  \frac{1}{\sqrt{k}}
\end{pmatrix}
\begin{pmatrix}
\cos (k\theta)  & -\sin (k\theta)  \\
\sin (k\theta)  & \cos  (k \theta) 
\end{pmatrix} 
\begin{pmatrix}
\frac{1}{\sqrt{k}} & 0 \\
0 & \sqrt{k}
\end{pmatrix}
\begin{pmatrix}
\e^w & 0 \\
0 & \e^{-w}
\end{pmatrix}
\label{susyInteractionForPositiveEnergy}
\end{equation}
or
\begin{equation}
A = 
\begin{pmatrix}
\sqrt{k}  & 0 \\
0 &  \frac{1}{\sqrt{k}}
\end{pmatrix}
\begin{pmatrix}
\cosh (k\theta)  & \sinh (k\theta)  \\
\sinh (k\theta)  & \cosh  (k \theta) 
\end{pmatrix} 
\begin{pmatrix}
\frac{1}{\sqrt{k}} & 0 \\
0 & \sqrt{k}
\end{pmatrix}
\begin{pmatrix}
\e^w & 0 \\
0 & \e^{-w}
\end{pmatrix}
\label{susyInteractionForNegativeEnergy}
\end{equation}
where $\theta \sim \text{Exp}(p)$ and $w$ is a random variable
independent of $\theta$. These products arise in the solution of the
generalised Kronig--Penney model 
with the supersymmetric interaction of Example \ref{supersymmetricExample}, i.e.
$$
B = \begin{pmatrix}
\e^{w} & 0 \\
0 & \e^{-w}
\end{pmatrix}\,.
$$

Let $\varrho$ denote the density of $w$. The Frisch--Lloyd equation
(\ref{frischLloydEquation}) is 
$$
  (z^2 + E)\,f(z) 
  + p \int_{\mathbb R}  \d w \int_{z}^{z \e^{-2w}}\d t \, f(t)\,\varrho(w) =N\,.
$$
After changing the order of integration, this becomes
\begin{equation}
  N = (z^2 + E)\,f(z) 
  + p \int_{0}^\infty  K\left( \frac{1}{2} \ln \frac{z}{t} \right)\,
  f(t) \,\d t
\label{frischLloydForSusy}
\end{equation}
where $K$ is the kernel defined by Equation (\ref{deltaKernel}).

Let
$$
\pm w \sim \text{\tt Exp}(q)\,.
$$
Then the kernel is supported on ${\mathbb R}_{\pm}$ and satisfies the
differential equation 
$$
K'(x) = \mp q K(x)\,, \;\; K(0\pm) = \mp 1\,.
$$
We 
deduce
\begin{equation}
  \frac{\d}{\d z} \left [ (z^2 + E) \,f(z) \right ] 
  - p \,f(z) \pm q \frac{ z^2 + E}{2z} \,f(z)  = \pm \frac{q}{2z}\, N
  \label{susyDifferentialEquation}
\end{equation}
where $N$ is the integrated density of states.  

Before going on to solve this equation, let us make a general remark: in the supersymmetric case, if one knows the invariant density, say
$f_+$, for a certain distribution 
of the strength $w$, then one can easily deduce the invariant density,
say $f_-$, when the sign of the strength is reversed. 
For instance, in the case $E=1$, the relationship between $f_-$ and
$f_+$ is simply 
$$
f_-(z) = \frac{1}{z^2} f_+ \left ( - \frac{1}{z} \right )\,.
$$
This relationship can be deduced directly from the form of the matrices in the product (\ref{susyInteractionForPositiveEnergy}).
It is also connected with the fact that changing the sign of the superpotential $W$ in Example \ref{supersymmetricExample} 
corresponds to swapping the functions
$\phi$ and $\psi$--- a manifestation of the so-called supersymmetry of the Hamiltonian. 

\subsubsection{Product of the form (\ref{susyInteractionForPositiveEnergy}).}
For $E=k^2 >0$ and $\pm w \sim {\tt Exp}(q)$, we have
\begin{multline}
  f(z) = N \frac{\mp \frac{q}{2} |z|^{ \mp \frac{q}{2}}}{z^2+k^2} 
   \exp \left[ \frac{p}{k} \arctan \frac{z}{k} \right] \\
   \times \int_z^{c_{\pm}(z)} | t |^{\pm \frac{q}{2}} 
   \exp \left[ -\frac{p}{k} \arctan \frac{t}{k} \right] \frac{\d t}{t}
  \label{susyExponentialDensityForEpositive}
\end{multline}
where
$$
  c_+(z) = 0 \;\;\text{and}\;\; c_{-}(z) = 
  \begin{cases} 
    \infty & \text{if $z>0$} \\
    -\infty & \text{if $z<0$}
  \end{cases}
  \,.
$$
 
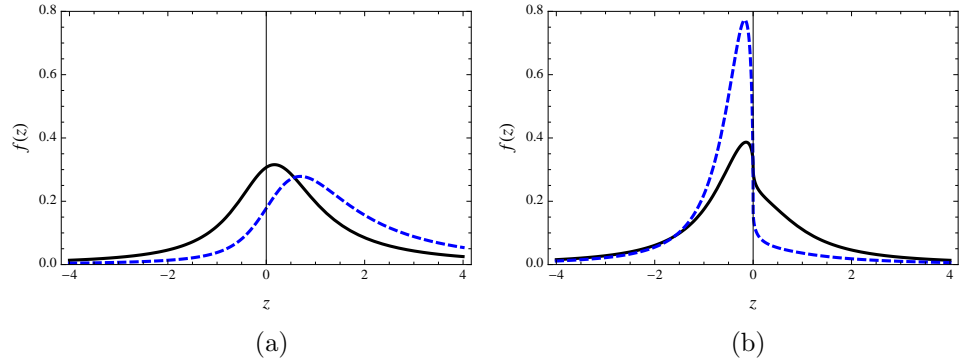
\begin{figure}[htbp]
  \centering
  \includegraphics[scale=0.725]{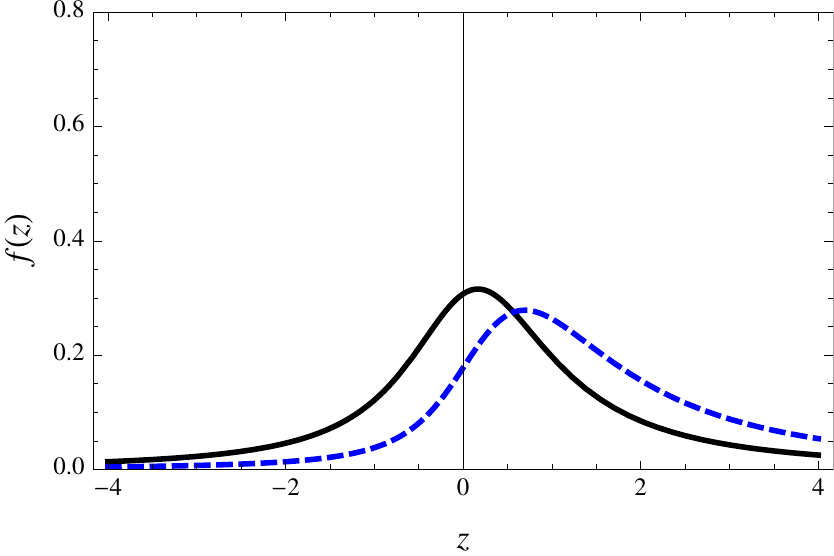}
  \hspace{0.1cm}
  \includegraphics[scale=0.725]{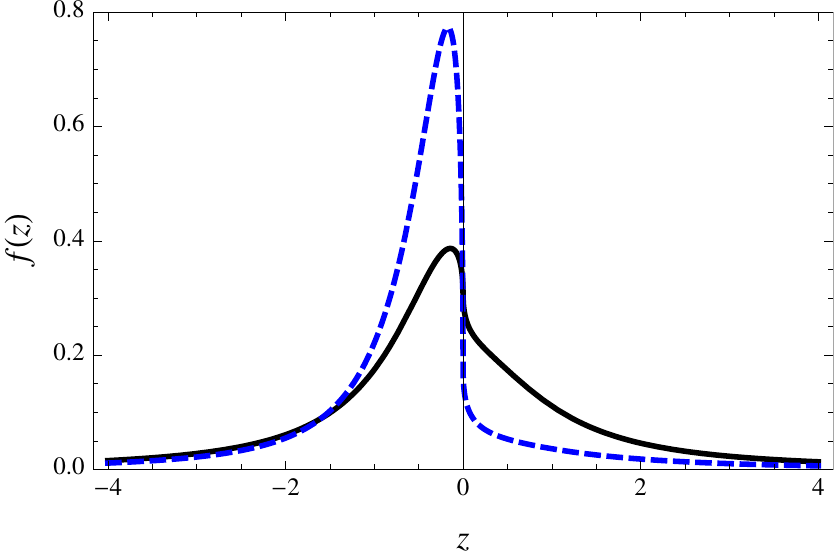}
  \vspace{0.1cm}
  \caption{Plots of the invariant density $f$ for supersymmetric
    scatterers and
    positive energy $E=k^2=+1$.
    Black continuous lines correspond to a ``low'' density of scatterers, i.e.
    $\theta\sim{\tt Exp}(p)$ with $p=1/2$, and blue dashed lines to a
    ``high'' density $p=2$.
   (a) $w\sim\text{\tt Exp}(1)$.
   (b) $-w\sim\text{\tt Exp}(1)$.}
  \begin{picture}(0,0) 
         \put(-85,80){(a)}
         \put(95,80){(b)}
  \end{picture}
  \label{fig:fsusy_pe}
\end{figure}

For the supersymmetric scatterer,
\begin{equation}
  {\mathscr B}(z) = \e^{2 w} z\,.
  \label{eq:Bsusy}
\end{equation}
Hence, for $w \sim {\tt Exp}(q)$, the jumps increase the Riccati
variable if it is already positive, and decrease it
otherwise. Furthermore there is no bound on the magnitude of the
jumps. It follows that the effect of increasing the density $p$ of the scatterers is to decrease
the density $f$ on ${\mathbb R}_-$, and to increase it on ${\mathbb R}_+$. This is in agreement with the plots shown in
Figure \ref{fig:fsusy_pe} (a). 

For $-w \sim {\tt Exp}(q)$, we observe the opposite effect: as shown in Figure \ref{fig:fsusy_pe} (b),
for increasing $p$,
the density $f$ is lowered on ${\mathbb R}_+$ and raised on ${\mathbb R}_-$. The asymmetry
of the plots for a negative strength $w$ is readily explained by using the particle analogy: starting at $+\infty$, the particle
rolls down the potential, spurred along by the impurities, and quickly
reaches the origin. Once the particle crosses over to the left, the impurities 
work against the downward force and tend to push the particle back towards the origin.

\subsubsection{Product of the form (\ref{susyInteractionForNegativeEnergy}).}
For $E = - k^2 <0$ and $w \sim {\tt Exp}(q)$,
we must take $N=0$ in Equation (\ref{susyDifferentialEquation}) to obtain
a normalisable solution. This is consistent with the well-known fact that the spectrum
of a supersymmetric Schr\"{o}dinger Hamiltonian must be contained in $\mathbb{R}_+$. Hence
\begin{equation}
  f(z) = C_+^{-1} \frac{z^{-\frac{q}{2}}}{z^2-k^2} 
  \left ( \frac{z-k}{z+k} \right )^{\frac{p}{2 k}} \,{\mathbf 1}_{(k,\infty)}(z)\,.
  \label{susyExponentialDensityForEnegativePlus}
\end{equation}
For $-w \sim {\tt Exp}(q)$, the solution is, instead,
\begin{equation}
  f(z) = C_-^{-1} \frac{z^{\frac{q}{2}}}{k^2-z^2} 
  \left ( \frac{k-z}{k+z} \right )^{\frac{p}{2 k}} \,{\mathbf 1}_{(0,k)}(z)\,.
  \label{susyExponentialDensityForEnegativeMinus}
\end{equation}
By Formula 8 in \cite{GR}, \S 3.197,
\begin{equation}
C_{\pm} = k^{\mp \frac{q}{2}-1} \, 
{\tt B}\!\left( \frac{q}{2}+1,\frac{p}{2k} \right) \,
{_2}F_1\!\left(
  \frac{p}{2k}+1,\frac{q}{2}+1;\frac{p}{2k}+\frac{q}{2}+1;-1 
\right)
\label{susyNormalisationConstant}
\end{equation}
where ${\tt B}$ is the beta function and ${_2}F_1$ is Gauss's
hypergeometric function.  

\begin{figure}[htbp]
  \centering
  \includegraphics[scale=0.725]{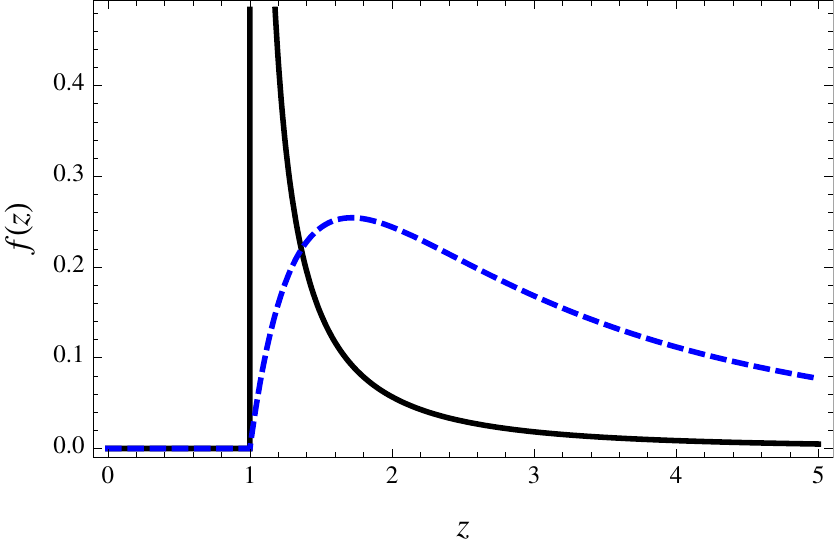}
  \hspace{0.1cm}
  \includegraphics[scale=0.725]{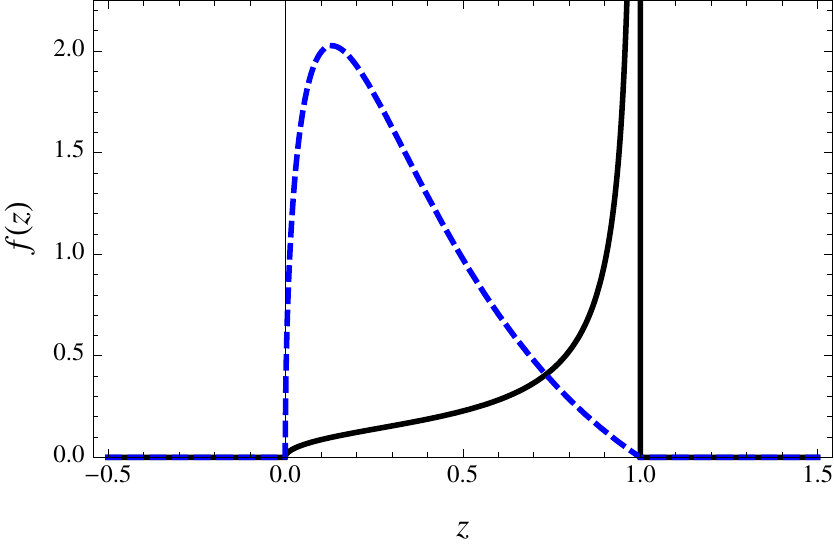}
  \vspace{0.1cm}
  \caption{Plots of the invariant density $f$  for supersymmetric
    scatterers and negative energy $E=-k^2=-1$.
    Black continuous lines correspond to a ``low'' density of scatterers, i.e.
    $\theta\sim{\tt Exp}(p)$ with $p=1/4$, and blue dashed lines to a
    ``high'' density $p=4$.
    (a) $w\sim\text{\tt Exp}(1)$.
   (b) $-w\sim\text{\tt Exp}(1)$.}
  \begin{picture}(0,0) 
         \put(-85,80){(a)}
         \put(95,80){(b)}
  \end{picture}
  \label{fig:fsusy_ne}
\end{figure}
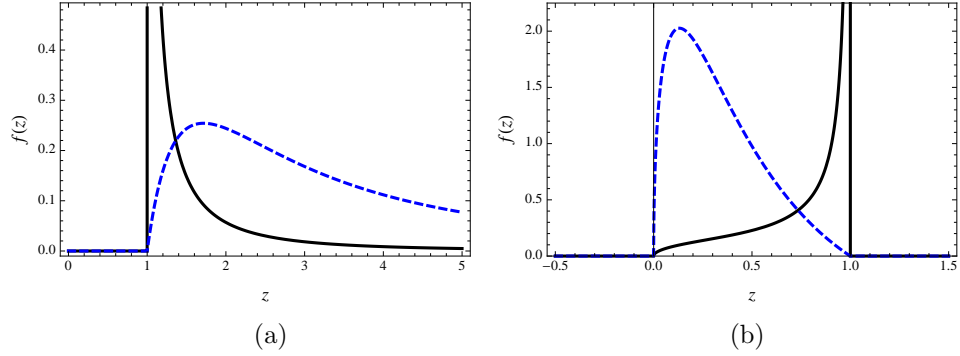

Plots of the invariant density are shown in Figure \ref{fig:fsusy_ne}.
As before, the particle analogy helps to explain their qualitative features: 
in view of Equation \eqref{eq:Bsusy}, when $w \sim {\tt Exp}(q)$,
the ``particle'' must eventually end 
up to the right of the equilibrium point $z=k$; see Figure \ref{fig:fsusy_ne} (b).
By contrast, when $-w \in {\tt Exp}(q)$ and $z>0$, the Riccati variable remains
positive but its value decreases at every jump. Hence, in this case, the
support of the invariant density is $(0,k)$; see Figure \ref{fig:fsusy_ne} (b).

\subsubsection{Calculation of the characteristic function.}
The essential spectrum--- and hence also the characteristic function--- is invariant under a change of sign of the strength $w$. 
So we need only consider the case
$w \sim {\tt Exp}(q)$.
For $E = -k^2$, we find, by using  Equation (\ref{susyExponentialDensityForEnegativePlus}),
\begin{equation}
\notag
\dashint_{-\infty}^{\infty} \d z \, z\, f(z)  = C_+^{-1} \int_k^{\infty} \d z \,
\frac{z^{-\frac{q}{2}+1}}{z^2-k^2} 
  \left ( \frac{z-k}{z+k} \right )^{\frac{p}{2 k}} \,.
\end{equation}
The normalisation constant $C_+$ is given explicitly by Formula
(\ref{susyNormalisationConstant}), and a similar formula 
is available for the definite integral; it suffices to replace $q/2$
by $q/2-1$. Using \eqref{reducedLyapunovForUpperTriangularB} and the
fact that $N(-k^2)=0$, the result is  
\begin{equation}
\notag
\Omega (-k^2) = -\frac p q +
 k \frac{{\tt B} \left ( \frac{p}{2 k}, \frac{q}{2} \right ) {_2} F_{1} \left ( \frac{p}{2k}+1, \frac{q}{2}; \frac{p}{2k}+\frac{q}{2};-1 \right ) }{{\tt B} \left ( \frac{p}{2 k}, \frac{q}{2} +1 \right ) {_2} F_{1} \left ( \frac{p}{2k}+1, \frac{q}{2}+1; \frac{p}{2k}+\frac{q}{2}+1;-1 \right )}\,.
\end{equation}
This formula extends to other values of the energy by analytic continuation; it suffices to replace $k$ by $\sqrt{-E}$.
In particular, for $E = k^2 > 0$, the characteristic function is obtained by replacing $k$ by $-\i k$. The density of states
and the Lyapunov exponent may then be deduced from the formulae (see Equation \eqref{characteristicFunction})
$$
  N = -\frac{1}{\pi} \,\text{Im} \,\Omega 
  \;\;\text{and}\;\; 
  \gamma = \text{Re} \,\Omega\,. 
$$


\section{Extensions}
\label{extensionSection}
In this final section, we consider possible extensions of our results: (1) to another scatterer; (2)
to another distribution of the strength of the scatterers and (3) to another distribution of the spacing between
consecutive scatterers.

\subsection{Double impurities}
\label{doubleImpuritySubsection}
The decomposition formula (\ref{gramSchmidtDecomposition}) gives a formal correspondence between products of $2 \times 2$ matrices
and generalised Kronig--Penney models of unit energy where the point scatterers are double impurities. In the particular case
$$
\theta \sim {\tt Exp} (p)\,,
$$
the density of the invariant measure solves the Frisch--Lloyd equation
\begin{equation}
N = (z^2+1) f(z) + p \int_{\mathbb R} \int_{\mathbb R} \d u\, \d w \, \varrho(u,w)\,\int_z^{z \e^{-2 w}-u} f(y)\,\d y \,.
\label{frischLloydForDoubleImpurity}
\end{equation}

We have already considered the cases where $w$ vanishes almost surely (the delta scatterer) or $u$ vanishes almost surely (the supersymmetric
scatterer). The purpose of this subsection is to consider the truly multivariate case where $u$ and $w$ are independent and
$$
u \sim {\tt Exp} (q_d)\,,\;\; w \sim {\tt Exp}(q_s)\,.
$$
We shall show that the corresponding invariant density $f$ solves the  differential
equation 
\begin{equation}
\frac{\d}{\d z} \left [ 2 z \left ( \varphi' - p \,f \right ) \right ] + \left ( q_s + 2 q_d z \right ) \left ( \varphi' - p\, f \right ) 
+ q_d q_s \varphi = q_d q_s N 
\label{doubleImpurityDifferentialEquation}
\end{equation}
where
$$
\varphi := (z^2+1) f(z)
$$
and $N$ is independent of $z$.
To derive this equation from (\ref{frischLloydForDoubleImpurity}), we shall consider the cases $z \le 0$ and $z>0$ separately.

Consider the latter case; 
we write
$$
\varrho (u,w) = \varrho_d (u) \,\varrho_s(w)\,.
$$ 
By changing
the order of integration, we find
\begin{multline}
\notag
\int_0^\infty \d w \,\varrho_s(w)\, \int_z^{z \e^{-2 w}-u} f(y) \,\d y \\
 =
-\int_{z-u}^{z} f(y)\,\d y + \int_{-u}^{z-u} K_s \left ( \frac{1}{2} \ln \frac{z}{y+u} \right ) f(y)\,\d y\,,
\end{multline}
where
$$
K_s (x) := \begin{cases}
-\e^{-q_s x} & \text{if $x \ge 0$} \\
0 & \text{if $x < 0$}
\end{cases}\,.
$$
So Equation (\ref{frischLloydForDoubleImpurity}) becomes
\begin{multline}
\notag
N = \varphi(z) - p \int_0^{\infty} \d u \,\varrho_d(u) \int_{z-u}^{z} f(y) \,\d y \\ +
p \int_0^{\infty} \d u \,\varrho_d(u) \int_{-u}^{z-u} K_s \left ( \frac{1}{2} \ln \frac{z}{y+u} \right ) f(y)\,\d y \,,
\end{multline}
and, by using integration by parts for the first integral on the right-hand side, we find
\begin{multline}
N = \varphi(z) - p \int_0^\infty \e^{-q_d u} f(z-u) \,\d u \\
+ p \int_0^{\infty} \d u \,\varrho_d(u) \int_{-u}^{z-u} K_s \left ( \frac{1}{2} \ln \frac{z}{y+u} \right ) f(y)\,\d y \,.
\end{multline}
Next, differentiate this equation with respect to $z$. By exploiting the identities
$$
K_s' = -q_s K_s\,,\;\; K_s(0+) = 1
$$
and
$$
\frac{\d}{\d u} \e^{-q_d u} = -q_d \,\e^{-q_d u}\,,
$$
we deduce
\begin{equation}
2 z \left [ \varphi'(z) - p \,f(z) \right ] + q_s \left [ \varphi(z)-N\right ] = q_s p \int_0^\infty \e^{-q_d u} \,f(z-u)\,\d u \,.
\end{equation}
The integral term may be eliminated by differentiating once more with respect to $z$, and we obtain eventually
Equation (\ref{doubleImpurityDifferentialEquation}).

The same equation is obtained if, instead, $z<0$. It is a trivial exercise to adapt these arguments to cater for cases
where one or both of $u$ and $w$ is always negative. 
We do not know how to express the solution of this second-order linear differential equation in terms of known
functions, except in the limiting cases
$$ 
q_d \;\text{fixed}\,,\;q_s \rightarrow \infty\,, 
$$
and
$$
q_d \rightarrow \infty\,,\; q_s \; \text{fixed}\,, 
$$
that have already been considered in \S \ref{deltaSubsection} and \S \ref{susySubsection} respectively.

\subsection{Delta scatterers with a gamma distribution}
\label{gammaSubsection}
The equation (\ref{frischLloydForDelta}) with the kernel (\ref{deltaKernel})
can be reduced to a purely differential form whenever $\varrho$ solves
a linear differential equation with piecewise constant coefficients. For instance,
suppose that
$$
\pm u \sim {\tt Gamma}(2,1/q)\,,
$$
i.e.
$$
\varrho (u) = 
\pm q^2 u \e^{\mp q u} {\mathbf 1}_{{\mathbb R}_{\pm}}(u)\,.
$$
Then
$$
\left ( \frac{\d}{\d u} \pm q \right )^2 \varrho = 0\,.
$$
Using the same trick as before, we obtain the following differential equation for $\varphi := (z^2+E) f$:
\begin{equation}
\left ( \frac{\d}{\d z} \pm q \right )^2 \varphi - p \left ( \frac{\d}{\d z} \pm 2 q \right ) \frac{\varphi}{z^2+E} = q^2 N\,.
\label{frischLloydGammaEquation}
\end{equation}
Suppose that $E = k^2 >0$ and use the ansatz
$$
\varphi (z) := \exp \left [  \mp q z + \frac{p}{k} \arctan \frac{z}{k} \right ] h (z)\,.
$$
Then
\begin{equation}
(z^2+k^2) \,h '' + p \,h' \mp p \,q \,h =0\,.
\label{gammaEquation}
\end{equation}
This equation may be solved in terms of hypergeometric series and, 
by imposing suitable auxiliary conditions, we can find a particular solution $h$ that is positive.  The method of variation of constants then yields
\begin{multline}
f(z) = \frac{q^2 N}{z^2+k^2} \exp \left [ \mp q z + \frac{p}{k} \arctan \frac{z}{k} \right ] h(z) \\
\times \int_{\mp \infty}^z
\exp \left [\pm q t - \frac{p}{k} \arctan \frac{t}{k} \right ]  H(t) \,\d t\,,
\label{densityForTheFrischLloydGammaCaseAndEpositive}
\end{multline}
where
$$
H(z) := \frac{\e^{\mp q z}}{h^2(z)} \int_{\mp \infty}^z \e^{\pm q t} h(t)\,\d t\,. 
$$

We now return to the calculation of the function $h(z)$ appearing in this formula. The general solution
of Equation (\ref{gammaEquation})
takes a remarkably simple form when
$$
p \,q = j(j-1)\,,\;\; j \in {\mathbb N}\,.
$$
Indeed, substitute
\begin{equation}
h(z) := \sum_{i=0}^\infty \frac{a_i}{i!} z^i\,,\;\; a_0 = 1\,,
\label{gammaSeries}
\end{equation}
into Equation (\ref{gammaEquation}). This yields a recurrence relation for the $a_i$:
$$
k^2 a_{i+2} = -p \,a_{i+1} + \left [ p\,q - i (i-1) \right ] a_{i-1} = 0\,,\;\; i=0,\,1,\, \ldots\,.
$$ 
By choosing $a_1$ so that $a_{j+1}=0$, the infinite series reduces to a polynomial, say $P_{j}$:
$$
P_2(z) = 1+2\,{\frac {pz}{{p}^{2}+2\,{k}^{2}}}+2\,{\frac {{z}^{2}}{{p}^{2}+2\,{
k}^{2}}}
\,.
$$
$$
P_3(z) = 1+6\,{\frac { \left( {p}^{2}+4\,{k}^{2} \right) z}{p \left( {p}^{2}+10
\,{k}^{2} \right) }}+18\,{\frac {{z}^{2}}{{p}^{2}+10\,{k}^{2}}}+24\,{
\frac {{z}^{3}}{p \left( {p}^{2}+10\,{k}^{2} \right) }}
\,.
$$
\begin{multline}
\notag
P_4(z) = 1+12\,{\frac {p \left( {p}^{2}+16\,{k}^{2} \right) z}{{p}^{4}+28\,{p}^
{2}{k}^{2}+72\,{k}^{4}}}+72\,{\frac { \left( {p}^{2}+6\,{k}^{2}
 \right) {z}^{2}}{{p}^{4}+28\,{p}^{2}{k}^{2}+72\,{k}^{4}}}\\
 +240\,{
\frac {p{z}^{3}}{{p}^{4}+28\,{p}^{2}{k}^{2}+72\,{k}^{4}}}+360\,{\frac 
{{z}^{4}}{{p}^{4}+28\,{p}^{2}{k}^{2}+72\,{k}^{4}}}
\,.
\end{multline}

Another solution  may be found by setting
$$
h(z) := (z^2+k^2) \exp \left [ -\frac{p}{k} \arctan \frac{z}{k} \right ] \sum_{i=0}^\infty \frac{b_i}{i!} \,z^i\,,\;\; b_0 = 1\,.
$$
Then 
$$
k^2 b_{i+2} = p \,b_{i+1} + \left [ p\,q - (i+2)(i+1) \right ] b_{i} = 0\,, \;\; i=0,\,1,\,\ldots\,.
$$
By choosing $b_1$ so that $b_{j-1} = 0$, this series reduces to another polynomial, say $Q_{j}$:
$$
Q_2 (z) = 1\,.
$$
$$
Q_3 (z) = 1- \frac{4}{p} z\,.
$$
$$
Q_4 (z) = 1-10\,{\frac {pz}{6\,{k}^{2}+{p}^{2}}}+30\,{\frac {{z}^{2}}{6\,{k}^{2}
+{p}^{2}}}\,.
$$
Hence the general solution of Equation (\ref{gammaEquation}) is
\begin{equation}
h(z) = c_1 P_j (z) + c_2  Q_j(z) (z^2+k^2) \exp \left [ -\frac{p}{k} \arctan \frac{z}{k} \right ]\,.
\label{generalSolution}
\end{equation}

Even with such detailed knowledge, it is not straightforward to identify the particular solution $h$ that yields the density. We end with the remark that
the characteristic function may, nevertheless, be constructed by using Halperin's trick: in this case, the homogeneous version of 
Equation (\ref{EquationForDeltaScatterers}) is
$$
F''(x) + \left \{ -E \pm \frac{p/q}{(1 \pm \i x/q)^2} \left [ 2 \pm \i x/q \right ] \right \} F(x) = 0\,.
$$
The solutions are expressible in terms of Whittaker functions; in particular, for $k$ real,
$$
\Omega (k^2) = \gamma (k^2) - \i \pi N (k^2) = 2\i k \,
    \frac{W_{-\i\frac{p}{2k},\frac{\sqrt{1 \pm 4pq}}{2}}' \left ( \mp 2 \i k q \right
    )}{W_{-\i\frac{p}{2k},\frac{\sqrt{1 \pm 4pq}}{2}} \left ( \mp 2 \i k q \right )}\,. 
$$
This result was originally found by Nieuwenhuizen \cite{Ni}.


\subsection{An alternative derivation of the Frisch--Lloyd equation}
\label{alternativeSubsection}
In deriving Equation (\ref{frischLloydEquation}), we made explicit use of the fact that,
when the spacing $\theta_j := x_{j+1}-x_j$ is exponentially distributed, 
$$
n(x) := \# \left \{ x_j :\; x_j < x \right \}
$$
is a Poisson process. In this subsection, we outline an alternative derivation of the Frisch--Lloyd
equation which generalises to other distributions of the $\theta_j$.

There is no real loss of generality in setting $E=1$. We use the decomposition
$$
A = \begin{pmatrix} \cos \theta & -\sin \theta \\ \sin \theta & \cos \theta \end{pmatrix}
\, B\,.
$$
Then
$$
{\mathscr A}^{-1} =  {\mathscr B}^{-1} \circ {\mathscr R}_{-\theta}
$$
where
$$
{\mathscr R}_{\theta} (z) := \frac{z \cos \theta-\sin \theta}{z \sin \theta+\cos \theta}\,.
$$
Denote by $\varrho$ the density of the random variable $\theta$ and by $\kappa$ the distribution of $B$.
Equation (\ref{integralEquation}) for the invariant density $f$ then becomes
\begin{multline}
\notag
f(z) = \int_{\mathbb R} \d \theta\, \varrho (\theta) \int_{\text{SL} \left ( 2,{\mathbb R} \right )}  \kappa ( \d B )
\left [ f \circ {\mathscr B}^{-1} \circ {\mathscr R}_{-\theta} \right ] (z) \frac{\d}{\d z} \left [ {\mathscr B}^{-1} \circ {\mathscr R}_{-\theta} \right ] (z) \\
= \int_{\mathbb R} \d \theta\, \varrho (\theta) \int_{\text{SL} \left ( 2,{\mathbb R} \right )}  \kappa ( \d B )
\left [ f \circ {\mathscr B}^{-1}  \right ] (w) \frac{\d {\mathscr B}^{-1}}{\d z} (w) \,\frac{\partial w}{\partial z}
\end{multline}
where 
$$
w := {\mathscr R}_{-\theta} (z) = \frac{z \cos \theta+ \sin \theta}{-z \sin \theta + \cos \theta}\,.
$$
The same equation can also be written in the more compact form
\begin{equation}
f(z) = \int_{\mathbb R} \d \theta\, \varrho (\theta) \frac{\partial}{\partial z} \int_{\text{SL} \left ( 2,{\mathbb R} \right )}
\kappa (\d B) \int_0^{{\mathscr B}^{-1}(w)} \d t f(t)\,.
\label{alternativeIntegralEquation}
\end{equation}
Now,
$$
\frac{\partial w}{\partial \theta} = (1+z^2) \frac{\partial w}{\partial z} \,. 
$$
Hence, if we multliply Equation (\ref{alternativeIntegralEquation}) by $1+z^2$, we obtain
$$
(1+z^2) f(z) = \int_{\mathbb R} \d \theta\, \varrho (\theta) \frac{\partial}{\partial \theta} \int_{\text{SL} \left ( 2,{\mathbb R} \right )}
\kappa (\d B) \int_0^{{\mathscr B}^{-1}(w)} \d t f(t)\,.
$$ 
Next, differentiate with respect to $z$:
\begin{multline}
\notag
\frac{\d}{\d z} \left [ (1+z^2) f(z) \right ] = \int_{\mathbb R} \d \theta\, \varrho (\theta) \frac{\partial^2}{\partial z \partial \theta} \int_{\text{SL} \left ( 2,{\mathbb R} \right )}
\kappa (\d B) \int_0^{{\mathscr B}^{-1}(w)} \d t f(t) \\
= \int_{\mathbb R} \d \theta\, \varrho (\theta) \frac{\partial^2}{\partial \theta \partial z} \int_{\text{SL} \left ( 2,{\mathbb R} \right )}
\kappa (\d B) \int_0^{{\mathscr B}^{-1}(w)} \d t f(t)\,.
\end{multline}
We may then use integration by parts for the outer integral; in the particular case
$$
\varrho (\theta) = p \e^{-p \theta} {\mathbf 1}_{[0,\infty)}
$$
the Frisch--Lloyd equation (\ref{frischLloydEquation}) follows easily after invoking (\ref{alternativeIntegralEquation}) once more.

We can use the same trick whenever the density of $\theta$ satisfies a linear differential equation with constant coefficients. For instance,
in the case
$$
\theta \sim {\tt Gamma} (2, p)
$$
it may be shown that
$$
\frac{\d}{\d z} \left [ (1+z^2) \varphi'(z) \right ] -2 p \varphi'(z) + p^2 f(z) =
p^2 \int_{\text{SL} \left (2,\mathbb R \right )} \kappa (\d B) \left [ f \circ {\mathscr B}^{-1} \right ](z) \frac{\d {\mathscr B}^{-1}}{\d z} (z) 
$$
where
$$
\varphi (z) = (1+z^2) f(z)\,.
$$


\section{Conclusion}
\label{sec:Conclusion}

In this article we have studied the invariant measure of products of
random matrices in $\text{SL}\left(2,{\mathbb R}\right)$.
This study relied on the correspondence between such products and a
certain class of random Schr\"odinger equations in which the potential
consists of point scatterers. 
We have considered several instances of this correspondence:
delta, delta-prime and supersymmetric scatterers.
By generalising the approach developed by Frisch \& Lloyd
for delta scatterers,
we have obtained an integral equation for the invariant density of a 
Riccati variable; this density yields the invariant measure of the
product of random matrices. 
For the three cases of point scatterers we have obtained
explicit formulae for the invariant measures. These are
the main new results of this paper.

The integrated density of states and the Lyapunov exponent of these
models were also calculated. Two approaches were used for this purpose:
the first is ``Halperin's trick'' and is specific to the
case of delta scatterers (cf. section~\ref{halperinSubsection});
the second uses analytic continuation of the characteristic function
and depends on the explicit knowledge of the invariant measure in some
interval of the energy outside the spectrum. 
By the first of these methods we have recovered the results of
Nieuwenhuizen in the case of delta scatterers.
By the second method
we have found new explicit formulae for the integrated density
of states and for the Lyapunov exponent in the cases of delta-prime
and of supersymmetric scatterers. 

All these analytical results were obtained when the spacing between
consecutive scatterers, as well as the impurity strength, have
exponential distributions.  
Possible extensions to the gamma distribution were also discussed.

A more complicated type of scatterer, combining the delta and the
supersymmetric scatterers, has also been examined. We called this
scatterer the ``double impurity''; it is interesting because every
product of matrices in $\text{SL}\left(2,{\mathbb R}\right)$ may in
principle be studied by considering a Schr\"odinger problem whose
potential consists of double impurities.
Although we succeeded in deriving a differential equation for the
invariant measure associated with a particular distribution of such
scatterers, we were unable to express its solution in terms of known
functions. 

In this paper we have played down the physical aspects of the models. 
Apart from the inverse localisation length and the density of states, 
there are other physical quantities that bear some relation to the
Riccati variable and whose statistical properties are of interest.
Let us mention three of them:
the most obvious is the phase of the reflexion coefficient on the 
disordered region; for a semi-infinite disordered region, its
distribution is trivially related to the invariant density of the
Riccati variable \cite{BarLuc90,GolNosSch94,Lu}.  
Another quantity is the Wigner time delay (the derivative of the phase
shift with respect to the energy); it has been considered in the
contexts of the Schr\"odinger \cite{JayVijKum89,TexCom99} and Dirac
\cite{SteCheFabGog99} equations.
A third quantity is the transmission coefficient (i.e. conductance)
\cite{AntPasSly81,DeyLisAlt01,SchTit03}.
The study of the distributions of the Wigner time delay and of the
transmission coefficient is mathematically more challenging because
it requires the analysis of some joint distributions; for this
reason it has been confined so far to limiting cases.

Some of the physical aspects arising from our results will form the basis of
future work.


\appendix


\section{Scattering, transfer and boundary matrices}
\label{app:ScatteringTransfer}

We discuss in this section the relationship between the scattering matrix $S$,
defined by \eqref{eq:ScatteringMatrix}, and
the boundary matrix $B$, defined by \eqref{pointInteraction}.
Here we need not assume that the scatterer is
necessarily pointlike; the scattering matrix and the corresponding boundary
matrix could equally well describe the effect of a potential supported on an interval.

We first write the scattering matrix in terms of transmission and reflexion
probability amplitudes $t$, $t'$ and $r$, $r'$~:
$$
  S = 
  \begin{pmatrix} r & t' \\ t & r' \end{pmatrix}
  \:.
$$
Current conservation implies
$$
|a^\text{out}_+|^2+|a^\text{out}_-|^2=|a^\text{in}_-|^2+|a^\text{in}_+|^2
$$ 
and so forces the scattering matrix to be unitary, i.e. $S \in U(2)$.
The constraints on the coefficients, namely
$$
|r|^2+|t|^2=|r'|^2+|t'|^2=|r'|^2+|t|^2=|r|^2+|t'|^2=1\,,
$$
$$
{r'}/{t'}=-\overline{{r}/{t}} \;\;\text{and}\;\; {r}/{t'}=-\overline{{r'}/{t}}\,,
$$
are conveniently built into
the following parametrisation, which also illustrates the factorisation $U(2)=U(1)\times{}SU(2)$:
\begin{equation}
  S = \i\e^{\i\theta}
  \begin{pmatrix}
    \e^{\i\varphi} \sqrt{1-\tau} &  -\i\e^{-\i\chi} \sqrt{\tau} \\[0.1cm]
    -\i\e^{\i\chi} \sqrt{\tau}   &     \e^{-\i\varphi} \sqrt{1-\tau} 
  \end{pmatrix}
  \:.
  \label{eq:ParametrisationS}
\end{equation}
This representation of the scattering matrix
is interesting because the four real parameters have a clear physical
interpretation:
$\tau\in[0,1]$ is the probability of transmission through the scatterer;
$\theta$ is the global phase of the matrix, i.e.
$$
\det{}S=-\e^{2\i\theta}\,.
$$
It is sometimes referred to as the ``Friedel phase'' 
since it is the phase appearing in the Krein--Friedel sum rule
relating the local density of states of the scattering region to a
scattering property.  
The phase $\varphi$
is a measure of the left-right asymmetry ($\varphi=0$ or $\pi$ corresponds to a
scattering invariant under $x\to-x$).
Finally the
phase $\chi$ is of magnetic origin, since time reversal corresponds to
transposition of the scattering matrix.

Next, we introduce the transfer matrix $T$ relating
left and
right amplitudes:
$$
\begin{pmatrix}
  a^\text{out}_+ \\[0.1cm] a^\text{in}_+
\end{pmatrix}
=
T
\begin{pmatrix}
  a^\text{in}_- \\[0.1cm] a^\text{out}_-
\end{pmatrix}
\hspace{0.5cm}\mbox{where}\hspace{0.5cm}
T = \begin{pmatrix}
  \overline{1/t} & -\overline{r/t} \\[0.1cm] -r/t' & 1/t' 
\end{pmatrix}
\:.
$$
This matrix is useful when considering the cumulative effect of many scatterers because it follows a simple composition law.
Again, current conservation implies that the transfer matrix is unitary:
$$
\left | a_+^{\text{out}} \right |^2-\left | a_+^{\text{in}} \right |^2 = \left | a_-^{\text{in}} \right |^2-\left | a_-^{\text{out}} \right |^2\,.
$$
In other words,
$T\in{}U(1,1)$ (note that $\det{}T=t/t'=\e^{2\i\chi}$).

The boundary matrix is also a
``transfer matrix'' in the sense that it connects properties of the
wavefunction on both sides of the scatterer. The relation between $T$ and $B$
is easily found:
from
$$
\begin{pmatrix}
  \psi'(0\pm) \\ \psi(0\pm)
\end{pmatrix}
=
\begin{pmatrix} \i k & -\i k \\ 1 & 1 \end{pmatrix}
\begin{pmatrix}
  a_{\pm}^{\substack{\text{out} \\ \text{in}}} \\[0.1cm] a_{\pm}^{\substack{\text{in} \\ \text{out}}}
\end{pmatrix}
$$
we deduce
$$
B=U\,T\,U^\dagger
\hspace{0.5cm}\mbox{where}\hspace{0.5cm}
U = \frac{\e^{-\i\pi/4}}{\sqrt{2k}}
\begin{pmatrix} \i k & -\i k \\ 1 & 1 \end{pmatrix}
\:.
$$
Then, using the parametrisation \eqref{eq:ParametrisationS}, we arrive at
the
following alternative form of
Equation \eqref{pointInteraction}:
\begin{equation}
  \label{eq:2}
  B = \frac{\e^{\i\chi}}{\sqrt{\tau}}
  \begin{pmatrix}
    \cos\theta - \sin\varphi\, \sqrt{1-\tau} 
  & -k \left [ \sin\theta + \cos\varphi\, \sqrt{1-\tau} \right ] \\[0.1cm]
    \frac1k \left [ \sin\theta - \cos\varphi\, \sqrt{1-\tau} \right ] 
  & \cos\theta + \sin\varphi\, \sqrt{1-\tau}
  \end{pmatrix}
\end{equation}
In particular, this expression shows clearly that
$$
\e^{-\i\chi}B\in\mathrm{SL}(2,{\mathbb R})\,.
$$
In one dimension, if a magnetic field is present, it may
always be removed by a gauge transformation. Furthermore, setting the
magnetic phase in the exponential factor to zero does not affect the
spectrum of the Schr\"{o}dinger operator. Hence there is no loss
of generality in restricting our attention to the case
$B\in\mathrm{SL}(2,{\mathbb R})$.

We end this appendix with some examples of scatterers, expressed
in terms of the parameters $\chi$, $\tau$, $\theta$ and $\varphi$.

\setcounter{example}{0}
\begin{example}
  For $\tau=1$, $\varphi=0$ and $\chi=0$, $B$ is the matrix describing a rotation of angle
  $\theta=k\ell$. In this case,  the ``scattering'' is equivalent to free propagation
  through an interval of length $\ell$.
\end{example}

\begin{example}
The scattering matrix for the delta impurity may be
    written as  
  $$
  S = \e^{\i\theta}
  \begin{pmatrix}
     \i\sin\theta & \cos\theta \\ \cos\theta & \i\sin\theta
  \end{pmatrix}
  $$
where
$$
\theta=-\arctan\frac{u}{2k} \in ( -\pi/2,\,\pi/2)\,.
$$
The other parameters are given by $\chi =0$,
$$
\varphi = \begin{cases}
0 & \text{if $u<0$} \\
\pi & \text{if $u>0$}
\end{cases}
$$
and
  $$
  \tau=\left [ 1+ \left ( \frac{u}{2k} \right )^2 \right ]^{-1}\,.
  $$
\end{example}

\begin{example}
  For the delta-prime scatterer, the scattering matrix $S$ has the
  same form as in the previous example, but this time with
  $$
    \theta=\arctan\frac{vk}{2} \in ( -\pi/2,\,\pi/2)\,,
  $$
  $$
    \chi = 0\,,
  $$
  $$
    \varphi = \begin{cases}
    0 & \text{if $v<0$} \\
    \pi & \text{if $v>0$}
    \end{cases}\,,
  $$
  and
  $$
    \tau=\left [1+\left ( \frac{vk}{2}\right )^2 \right ]^{-1}\,.
  $$
\end{example}

\begin{example}
  The supersymmetric scatterer corresponds to taking $\chi=\theta=0$,
  $\varphi=-\pi/2$ and $\tau=\text{sech}^2 w$.
\end{example}

\bibliographystyle{amsplain}

\end{document}